\documentclass[12pt]{article}
\usepackage{amssymb}
\usepackage{graphicx}
\usepackage{multirow}
\usepackage{float}
\usepackage{dsfont}
\usepackage{placeins}
\usepackage{comment}
\usepackage{mathtools}
\usepackage{enumitem}
\usepackage{caption,setspace}
\usepackage{amsmath,amsthm}
\usepackage{longtable}
\graphicspath{{Plots/}}
\newcommand{\rgld}{RS G$\lambda$D }
\newcommand{\rglds}{RS G$\lambda$D}
\newcommand{\zrgld}{HRS G$\lambda$D }
\newcommand{\zrglds}{HRS G$\lambda$D}
\newcommand{\fgld}{FKML G$\lambda$D }
\newcommand{\fglds}{FKML G$\lambda$D}
\newcommand{\zfgld}{HFKML G$\lambda$D }
\newcommand{\zfglds}{HFKML G$\lambda$D}
\newcommand{\gld}{G$\lambda$D }
\newcommand{\glds}{G$\lambda$D}
\newcommand{\zgld}{HG$\lambda$D }
\newcommand{\zglds}{HG$\lambda$D}
\newcommand{\ml}{NMLM }
\newcommand{\mls}{NMLM}
\newcommand{\gpd}{GPD }
\newcommand{\gpds}{GPD}

\usepackage[natbibapa]{apacite}
\begin{document}
\centerline{ARTICLE}
\vskip 3mm

\noindent A SURVEY OF A HURDLE MODEL FOR HEAVY-TAILED DATA BASED ON THE GENERALIZED LAMBDA DISTRIBUTION
\vskip 3mm

\vskip 5mm
\noindent D. Marcondes$^{\dagger}$, C. Peixoto$^{\dagger}$ and A. C. Maia$^{\ddagger}$

\noindent $^{\dagger}$ Instituto de Matem\'atica e Estat\'istica, Universidade de S\~ao Paulo, Brazil

\noindent $^{\ddagger}$ Faculdade de Economia, Administra\c{c}ç\~ao e Contabilidade, Universidade de S\~ao Paulo, Brazil

\noindent dmarcondes@ime.usp.br

\vskip 3mm

\noindent ABSTRACT

In this survey we present an extensive research of the vast literature about the Generalized Lambda Distribution (\glds) and propose a hurdle, or two-way, model whose associated distribution is the \gld in order to meet the demand for a highly flexible model of heavy-tailed data with excess of zeros. We apply the developed models to a dataset consisting of yearly healthcare expenses, a typical example of heavy-tailed data with excess of zeros. The fitted models are compared with models based on the Generalised Pareto Distribution and it is established that the \gld models perform best.
\vskip 4mm

\noindent 1. INTRODUCTION

A motivation for the development of models for heavy-tailed data with excess of zeros arises from data on healthcare expenses, that is characterized by its heavy tails, its great number of zeros and its high skewness, which makes fitting models to it a complex task \citep{mihaylova2011,jones2014}. Indeed, a suitable choice of model for healthcare expenses are clumped-at-zero models, that are those with excess of zeros. The clumped-at-zero models are divided into two classes, as follows. The first class are the zero-inflated models, which are based on distributions that already have a probability mass at zero, that is then inflated. The zero-inflated Poisson model is an element of this class \citep{lambert1992}. The second class of clumped-at-zero models are the two-part or hurdle models, that are those whose underlying distribution does not have a probability mass at zero, that is then added to it. They are called hurdle for the probability mass at zero may be seem as a \textit{hurdle}. In the same sense, they are also known as two-part models because the probability mass at zero and the non-zero values may be modelled independently of each other, i.e., the model has two parts. An example of hurdle model, for the demand of medical care, is presented in \citet{duan1983}. In the class of two-part models there are also models whose underlying distribution has a probability mass at zero, but are nonetheless two-part models, as the model of \citet{mullahy1986}, since the inflation of the probability mass at zero is made independently of the non-zero data by truncation. 

The underlying distribution of the hurdle model treated in this paper is the Generalized Lambda Distribution (\glds), that is a highly flexible four-parameter continuous probability distribution. This distribution was first proposed by \citet{RS1974}, and then extended by \citet{FKML1988}, as a generalization of Tukey's Lambda Distribution \citep{hastings1947,Tukey1960}. Even though the \gld is a \textit{wild card} distribution, that well approximate others \citep[Chapter~3]{karian2000}, its use has been limited in the literature as there is no explicit expression for its probability density function, which makes it a complex task to estimate its parameters. 

Indeed, the estimation of the parameters of the \gld had been carried out by the methods of moments and a percentile method until \citet{su2007} proposed a numerical maximum likelihood method for it. Another limitation for the use of the \gld was the lack of a regression model, that was just recently proposed by \citet{su2015}, which extended the range of applications for the \glds. Therefore, due to recent advances in the theory of the \glds, it is now possible to further apply this powerful distribution and compare it to other established models in order to assess its advantages.

Although the estimation techniques for the \gld have been limited, there is a considerable amount of applications of it in the literature. As examples, we cite the evaluation of non-normal process capability indices \citep{pal2004}, option pricing \citep{corrado2001}, the fitting of solar radiation data \citep{ozturk1982} and income data \citep{tarsitano2004}, and statistical process control \citep{fournier2006}. Regarding the modelling of healthcare expenses, the \gld was studied by \citet{balasooriya2008}, where it was compared with the transformed kernel density and models of the exponential family, and it was established that the \gld fitted the data best. 

In this paper, we develop hurdle \gld models and assess their goodness-of-fit on a yearly healthcare expenses dataset. The models developed seek to fit the data taking into account covariates (regression model) or not. The \gld models are compared with hurdle models based on the Generalized Pareto Distribution (\gpds), that are special cases of the model in \citet{couturier2010}. The \gpd is also a highly flexible continuous probability distribution, although we argue that it is not as flexible as, and do not fit the data as good as, the \glds. For an assessment of the goodness-of-fit of the \gpd for healthcare expenses see \citet{cebrian2003}.

In Section 2 we present a survey about the \gld and its estimation techniques. In Section 3 we propose a hurdle \gld and develop its main properties. In Section 4 we present a survey about \gld regression models, and develop a hurdle \gld regression model. In Section 5 we present a simulation study about the asymptotic properties of the hurdle \gld regression coefficients. In Section 6 we apply the developed methods to model healthcare expenses and compare \gld models and \gpd models.
\vskip 4mm

\noindent 2. THE GENERALIZED LAMBDA DISTRIBUTION

In this section we present two distinct parametrizations of the \glds, known as the RS and \fglds, and some of their properties. 
\vskip 4mm

\noindent 2.1 RS GENERALIZED LAMBDA DISTRIBUTION

The \rglds, as proposed by \citet{RS1974}, is a four parameter generalization of Tukey's Lambda Distribution, obtained from an uniform random variable. Let $U$ be an uniform random variable with range $[0,1]$ defined in a probability space $(\Omega,\mathcal{F},\mathbb{P})$. Then, the random variable $X_{\boldsymbol{\lambda}}$, also defined in $(\Omega,\mathcal{F},\mathbb{P})$, and given by
\vspace{-0.5cm}
\begin{equation}
\label{RS}
X_{\boldsymbol{\lambda}} \coloneqq  Q_{\boldsymbol{\lambda}}(U) = \lambda_{1} + \frac{U^{\lambda_{3}} - (1 - U)^{\lambda_{4}}}{\lambda_{2}}
\vspace{-0.5cm}
\end{equation}
has an \rgld with parameters $\boldsymbol{\lambda} = (\lambda_{1},\lambda_{2},\lambda_{3},\lambda_{4})$. The function $Q_{\boldsymbol{\lambda}}(u), u \in [0,1]$, is the quantile function of $X_{\boldsymbol{\lambda}}$ as $Q_{\boldsymbol{\lambda}}(u) = F_{\boldsymbol{\lambda}}^{-1}(u)$, in which $F_{\boldsymbol{\lambda}}(x) = \mathbb{P}(X_{\boldsymbol{\lambda}} \leq x)$. The density of $X_{\boldsymbol{\lambda}}$ is given by
\vspace{-0.5cm}
\begin{equation}
\label{RSdensity}
f_{\boldsymbol{\lambda}}(x) = \frac{1}{Q'_{\boldsymbol{\lambda}}(F_{\boldsymbol{\lambda}}(x))} = \frac{\lambda_{2}}{\lambda_{3} F_{\boldsymbol{\lambda}}(x)^{\lambda_{3} - 1} + \lambda_{4} (1 - F_{\boldsymbol{\lambda}}(x))^{\lambda_{4}-1}}
\vspace{-0.5cm}
\end{equation}
in which $Q'_{\boldsymbol{\lambda}}(F_{\boldsymbol{\lambda}}(x))$ is the derivative of $Q_{\boldsymbol{\lambda}}$ at point $F_{\boldsymbol{\lambda}}(x)$. The parametric space $\Lambda = \{\boldsymbol{\lambda} \in \mathbb{R}^{4}: F_{\boldsymbol{\lambda}}$ is a cumulative distribution function$\}$ of $\boldsymbol{\lambda}$ is a proper subset of $\mathbb{R}^{4}$ and is given implicitly by inequality 
\vspace{-0.5cm}
\begin{equation}
\label{inequalityRS}
\frac{\lambda_{3} u^{\lambda_{3} - 1} + \lambda_{4} (1 - u)^{\lambda_{4}-1}}{\lambda_{2}} \geq 0
\vspace{-0.5cm}
\end{equation}
for $0 \leq u \leq 1$. The inequality is obtained noting that $f_{\boldsymbol{\lambda}}(x) \geq 0$ if, and only if, $Q'_{\boldsymbol{\lambda}}(F_{\boldsymbol{\lambda}}(x)) \geq 0$.

The \rgld is quite flexible, as it is possible to specify its parameters in order to obtain a specific distribution with given mean, variance, skewness and kurtosis. Indeed, the mean can be shifted to any value by choosing $\lambda_{1}$ properly, the skewness and kurtosis are determined by $\lambda_{3}$ and $\lambda_{4}$ and, given $\lambda_{3}$ and $\lambda_{4}$, the variance is determined by $\lambda_{2}$. The range of $X_{\boldsymbol{\lambda}}$ is $[Q_{\boldsymbol{\lambda}}(0),Q_{\boldsymbol{\lambda}}(1)]$ and depends on $\boldsymbol{\lambda}$ (see \citet[Theorem~1.4.23]{karian2000} for the \rgld range). The \textit{kth} moment of the \rgld exists if, and only if, $\min(\lambda_{3},\lambda_{4}) > -k^{-1}$ and, when it exists and $\lambda_{1} = 0$, it is given by
\vspace{-0.5cm}
\begin{equation}
\label{momentRS}
E(X_{\boldsymbol{\lambda}}^{k}) = \lambda_{2}^{-k}\sum_{i=0}^{k} \binom{k}{i} (-1)^{i} \beta(\lambda_{3}(k-i) + 1;\lambda_{4}i+1)
\vspace{-0.5cm}
\end{equation}
in which $\beta(a,b)$ is the beta function evaluated at $(a,b)$. A proof for (\ref{momentRS}) is given in \citet{RS1974}. The central moments of $X_{\boldsymbol{\lambda}}$ when $\lambda_{1} \neq 0$ may be obtained from (\ref{momentRS}) by applying the properties of the expectation operator. For instance, we have that 
\vspace{-0.5cm}
\begin{equation}
\label{meanRS}
E(X_{\boldsymbol{\lambda}}) = \lambda_{1} +\frac{(\lambda_{3} + 1)^{-1} - (\lambda_{4} + 1)^{-1}}{\lambda_{2}}
\vspace{-0.5cm}
\end{equation}
so that $E(X_{\boldsymbol{\lambda}}) = \lambda_{1}$ if, and only if, $\lambda_{3} = \lambda_{4}$ and $X_{\boldsymbol{\lambda}}$ is symmetric.

The estimation of the \rgld parameters may be performed by various methods. The classical estimation technique is the Method of Moments (MM), as introduced by \citet{RS1974} and consolidated by \citet{karian1996}. Although easily implemented nowadays, the MM has some limitations. First of all, two different vectors $\boldsymbol{\lambda_{1}}, \boldsymbol{\lambda_{2}} \in \Lambda$ may yield the same first four moments of the \rglds. As pointed out by \citet{karian1996} it may be seen as a problem or an opportunity, for it enables a flexible fit for the data, as we may choose the parameters that best fulfil our objectives regarding the fit. Another limitation of the MM is the fact that the existence of the first four moments depends on $\boldsymbol{\lambda}$ and, therefore, it cannot be applied for a subset of $\Lambda$. Furthermore, simulation studies have showed that the MM performs worse than other methods, as the Numerical Maximum Likelihood Method (\mls) and the percentile matching approach, for example \citep{karian2003,su2007}.

Even though other methods, as the \textit{least square} estimation method proposed by \citet{ozturk1985}, the Starship Method developed by \citet{king1999}, the flexible discretized approach proposed by \citet{su2005} and the percentile matching approach, similar to the MM but with best results in simulation studies, as introduced by \citet{karian1999} and further studied by \citet{karian2000} and \citet{karian2003}, are available in the literature, this paper treats only estimation by the \mls, as proposed by \citet{su2007} and \citet{su2011}. For a good account of other estimation techniques see \citet{lakhany2000}.

The log-likelihood of a sample $\{x_{1},\dots,x_{n}\}$ of an \rgld random variable may be written in terms of the cumulative distribution function $F_{\boldsymbol{\lambda}}$, by denoting $u_{i} = F_{\boldsymbol{\lambda}}(x_{i}), i = 1, \dots, n$, so that
\vspace{-0.1cm}
\begin{equation}
\label{loglRS}
l_{RS}(\boldsymbol{\lambda}) = \sum_{i=1}^{n} \log \Bigg[  \frac{\lambda_{2}}{\lambda_{3} u_{i}^{\lambda_{3} - 1} + \lambda_{4} (1 - u_{i})^{\lambda_{4}-1}}\Bigg], \boldsymbol{\lambda} \in \Lambda.
\vspace{-0.1cm}
\end{equation}
In order to maximize (\ref{loglRS}) it is preferable to apply direct numerical methods than the usual method of differentiation, as they are much more reliable and efficient than solving the conventional linear equations on $\boldsymbol{\lambda}$, because, in many cases, the \rgld may be undefined for certain parameters values, as was pointed out by \citet{su2011}. Therefore, we apply the algorithm proposed by \citet{su2007} to maximize (\ref{loglRS}). 

The main issue in maximizing (\ref{loglRS}) is in finding suitable initial values for the quantile sample $\{u_{1},\dots,u_{n}\}$. The most efficient way of obtaining initial values for them is through the estimation of $\boldsymbol{\lambda}$ by the percentile method, as this is the method that, apart from the \mls, has had more efficient results estimating the \rgld parameters \citep{karian2003}. The percentile method, as presented in \citet{karian2000} and \citet{su2007}, is as follows. The p\textit{th} percentile of a sample $\{x_{1},\dots,x_{n}\}$ is defined as $\hat{\pi}_{p} = x_{(r)} + k(x_{(r+1)} - x_{(r)})$, in which $\{x_{(1)},\dots,x_{(n)}\}$ is the sample ordered in ascending order and $r$ is the greatest integer lesser than $(n+1)p$, with $k = (n+1)p - r$. Rather than matching the sample moments to their theoretical value, in the percentile method we match the statistics
\vspace{-0.5cm}
\begin{align}
\label{matchPM}
\hat{\rho}_{1} = & \hat{\pi}_{0.5} & \hat{\rho}_{2} = & \hat{\pi}_{1-v} - \hat{\pi}_{v} &
\hat{\rho}_{3} = & \frac{\hat{\pi}_{0.5} - \hat{\pi}_{v}}{\hat{\pi}_{1-v} - \hat{\pi}_{0.5}} &
\hat{\rho}_{4} = & \frac{\hat{\pi}_{0.75} - \hat{\pi}_{0.25}}{\hat{\rho}_{2}}
\end{align}
to their theoretical values, in which $v$ is an arbitrary number between $0$ and $0.25$, that we choose to be $0.1$, so that it is consistent with \citet{karian2000} and \citet{su2007}.

Matching the theoretical values of $\rho_{1}, \rho_{2}, \rho_{3}$ and $\rho_{4}$ to the quantile function of an \rgld we obtain the following relations between $\rho_{1}, \rho_{2}, \rho_{3}, \rho_{4}$ and $\boldsymbol{\lambda}$:
\vspace{-0.5cm}
\begin{align}
\label{relationRS}
\rho_{1}(\boldsymbol{\lambda}) = & Q_{\boldsymbol{\lambda}}(0.5) = \lambda_{1} + \frac{0.5^{\lambda_{3}} - 0.5^{\lambda_{4}}}{\lambda_{2}} \nonumber \\
\rho_{2}(\boldsymbol{\lambda}) = & Q_{\boldsymbol{\lambda}}(1-v) - Q_{\boldsymbol{\lambda}}(v) = \frac{(1-v)^{\lambda_{3}} - v^{\lambda_{3}} + (1-v)^{\lambda_{4}} - v^{\lambda_{4}}}{\lambda_{2}} \nonumber \\
\rho_{3}(\boldsymbol{\lambda}) = & \frac{Q_{\boldsymbol{\lambda}}(0.5) - Q_{\boldsymbol{\lambda}}(v)}{Q_{\boldsymbol{\lambda}}(1-v) - Q_{\boldsymbol{\lambda}}(0.5)} = \frac{(1-v)^{\lambda_{4}} - v^{\lambda_{3}} + 0.5^{\lambda_{3}} - 0.5^{\lambda_{4}}}{(1-v)^{\lambda_{3}} - v^{\lambda_{4}} + 0.5^{\lambda_{4}} - 0.5^{\lambda_{3}}} \nonumber \\
\rho_{4}(\boldsymbol{\lambda}) = & \frac{Q_{\boldsymbol{\lambda}}(0.75) - Q_{\boldsymbol{\lambda}}(0.25)}{\rho_{2}} = \frac{0.75^{\lambda_{3}} - 0.25^{\lambda_{4}} + 0.75^{\lambda_{4}} - 0.25^{\lambda_{3}}}{\rho_{2}}.
\vspace{-0.5cm}
\end{align}
The conditions $- \infty < \rho_{1} < \infty$, $\rho_{2} \geq 0$, $\rho_{3} \geq 0$ and $\rho_{4} \in [0,1]$ must be satisfied, as can be established from (\ref{matchPM}). In order to estimate $\boldsymbol{\lambda}$ we match the sample values (\ref{matchPM}) to their theoretical values (\ref{relationRS}) and solve numerically for $\boldsymbol{\lambda}$ by the Newton-Raphson method, for example, with the stopping rule given by the minimization of the Euclidean 2-norm $H(\boldsymbol{\lambda}) = \lVert (\rho_{3}(\boldsymbol{\lambda}),\rho_{4}(\boldsymbol{\lambda})) - (\hat{\rho}_{3},\hat{\rho}_{4}) \rVert_{2}$. Once $\lambda_{3}$ and $\lambda_{4}$ are obtained from the last two equations of (\ref{relationRS}), we may substitute their values in the first two equations of (\ref{relationRS}) in order to obtain $\lambda_{1}$ and $\lambda_{2}$.

The percentile method is applied to get initial values in order to maximize (\ref{loglRS}). The maximization of (\ref{loglRS}) is performed by a 4-step algorithm proposed by \citet{su2007}\footnote{The algorithm in \citet{su2007} has five steps, that we reduced to four, without loss of content.}, that uses quasi random numbers and the percentile method. The algorithm is as follows:

\begin{enumerate}[nolistsep]	
	\item Specify the range of initial values for $\lambda_{3}$ and $\lambda_{4}$ and the number of values to be selected. In this step, quasi random numbers are sampled as candidates for the initial values of $\lambda_{3}$ and $\lambda_{4}$. \citet{su2007} proposes that $10,000$ quasi random values (scrambled so that the sampled values fill uniformly the considered space) be chosen from the square $[-1.5,1.5]^{2}$.
	\item Evaluate $\lambda_{1}, \lambda_{2}$ for each of the initial values of $\lambda_{3}, \lambda_{4}$ in the first two equations of (\ref{relationRS}). Remove all initial values that 
	\begin{enumerate}
		\item[(a)] Do not result in a legal parametrization of the \rgld by (\ref{inequalityRS}).
		\item[(b)] Do not span the entire region of the dataset.
	\end{enumerate}
	Among the initial points not excluded by step $2$, find the initial set $\hat{\boldsymbol{\lambda}}_{0}$ that minimizes the norm $H(\boldsymbol{\lambda})$.
	\item Calculate the quantiles $\{u_{1},\dots,u_{n}\}$ by solving numerically (\ref{RS}) with the initial values $\hat{\boldsymbol{\lambda}}_{0}$.
	\item Once $\{u_{1},\dots,u_{n}\}$ is obtained, substitute them in (\ref{loglRS}) and solve it numerically for $\hat{\boldsymbol{\lambda}}$. It is convenient to repeat this process for different initials values, in order to check the consistency of the solution. The obtained estimator is called revised percentile estimator of the \rgld under maximum likelihood estimation. The quality of the final fitting may be established by diagnostic techniques, as the data histogram superimposed by the estimated density, quantile plots and goodness-of-fit tests.
\end{enumerate} 

\vskip 4mm

\noindent 2.2 FKML GENERALIZED LAMBDA DISTRIBUTION

The \fglds, as proposed by \citet{FKML1988}, is also a four parameter generalization of Tukey's Lambda Distribution obtained from an uniform distribution. Indeed, let $U$ be an uniform random variable with range $[0,1]$ defined in a probability space $(\Omega,\mathcal{F},\mathbb{P})$. Then, the random variable $X_{\boldsymbol{\lambda}}$, also defined in $(\Omega,\mathcal{F},\mathbb{P})$, and given by
\vspace{-0.5cm}
\begin{equation}
\label{FKML}
X_{\boldsymbol{\lambda}} \coloneqq Q_{\boldsymbol{\lambda}}(U) = \lambda_{1} + \frac{1}{\lambda_{2}} \Bigg[\frac{U^{\lambda_{3}} - 1}{\lambda_{3}} - \frac{(1 - U)^{\lambda_{4}} - 1}{\lambda_{4}}\Bigg]
\vspace{-0.5cm}
\end{equation}
has an \fgld with parameters $\boldsymbol{\lambda} = (\lambda_{1},\lambda_{2},\lambda_{3},\lambda_{4})$. The \fgld is a probability distribution for all real-valued parameters $\boldsymbol{\lambda}$, with the restriction that $\lambda_{2} > 0$ and the conventions that $X_{(\lambda_{1},\lambda_{2},0,\lambda_{4})} = \lim\limits_{\lambda_{3} \rightarrow 0} X_{(\lambda_{1},\lambda_{2},\lambda_{3},\lambda_{4})}$ and $X_{(\lambda_{1},\lambda_{2},\lambda_{3},0)} = \lim\limits_{\lambda_{4} \rightarrow 0} X_{(\lambda_{1},\lambda_{2},\lambda_{3},\lambda_{4})}$. The main motivation for generalizing Tukey's Lambda distribution to (\ref{FKML}) is the weaker restrictions on its parametric space when comparing to the \rglds, which facilitates the estimation of its parameters. Although both the RS and \fgld are generalizations of Tukey's Lambda Distribution, they are not equivalent, so the distribution fitted by one parametrization to a dataset differs in general from the one fitted by the other. 

The range of $X_{\boldsymbol{\lambda}}$ is dependent on the parameters $\boldsymbol{\lambda}$ and is given by $[Q_{\boldsymbol{\lambda}}(0),Q_{\boldsymbol{\lambda}}(1)]$. The density of the \fgld is obtained in a similar manner of (\ref{RSdensity}) and is given by
\vspace{-0.1cm}
\begin{equation}
\label{FKMLdensity}
f_{\boldsymbol{\lambda}}(x) = \frac{1}{Q'_{\boldsymbol{\lambda}}(F_{\boldsymbol{\lambda}}(x))} = \frac{\lambda_{2}}{F_{\boldsymbol{\lambda}}(x)^{\lambda_{3} - 1} + (1 - F_{\boldsymbol{\lambda}}(x))^{\lambda_{4}-1}}.
\vspace{-0.1cm}
\end{equation}
The distribution of $X_{\boldsymbol{\lambda}}$ is symmetric if, and only if, $\lambda_{3} = \lambda_{4}$, although its skewness measure may be zero for\footnote{This is also the case for the \rglds.} $\lambda_{3} \neq \lambda_{4}$. The parameters $\lambda_{3}$ and $\lambda_{4}$ determine single-handedly the nature and shape of the left and right tails of $X_{\boldsymbol{\lambda}}$, respectively, although the shape of the probability density function depends on both $\lambda_{3}$ and $\lambda_{4}$. Examples of \fgld may be found in \citet{su2015}. Although the parameters of both the RS and \fgld are denoted by $\lambda_{1}$, $\lambda_{2}$, $\lambda_{3}$ and $\lambda_{4}$, and are related to the same properties of the distribution, they are not equivalent, nor comparable. 

The \textit{kth} moment of the \fgld also exists if, and only if, $\min(\lambda_{3},\lambda_{4}) > -k^{-1}$. Making $a = 1/\lambda_{2}$ and $b = \lambda_{1} - 1/\lambda_{2}\lambda_{3} + 1/\lambda_{2}\lambda_{4}$, the \textit{kth} moment of $X_{\boldsymbol{\lambda}}$ may be obtained from the moments of $(X_{\boldsymbol{\lambda}} - b)/a$ that, when exist, are given by
\vspace{-0.5cm}
\begin{equation}
\label{momentFKML}
s_{k} \coloneqq E\Bigg(\bigg[\frac{X_{\boldsymbol{\lambda}} - b}{a}\bigg]^{k}\Bigg) = \sum_{i=0}^{k} \binom{k}{i} (-1)^{i} \lambda_{3}^{-(k-i)} \lambda_{4}^{-i} \beta(\lambda_{3}(k-i)+1;\lambda_{4}i+1)
\vspace{-0.5cm}
\end{equation}
as showed in \citet{FKML1988} and \citet{lakhany2000}. The central moments of $X_{\boldsymbol{\lambda}}$ may also be obtained from (\ref{momentFKML}).

The \fgld is also highly flexible, as it is possible to choose $\boldsymbol{\lambda}$ so that $X_{\boldsymbol{\lambda}}$ has specific mean, variance, skewness and kurtosis. Furthermore, its tails are also flexible, so that the \fgld (and the \rglds) provides a better fit for heavy tailed data than the usual Generalized Additive Models for Location, Scale and Shape \citep{gamlss}, for example. However, the \fgld probability density function does not have an analytic form that does not depend on $F_{\boldsymbol{\lambda}}$, what calls for computational tools in order to fit it to a dataset.

Although there is also a vast literature about the estimation of the \fgld parameters, we treat only the \ml as proposed by \citet{su2007} and \citet{su2011}. The log-likelihood of a sample $\{x_{1},\dots,x_{n}\}$ of an \fgld is given by
\vspace{-0.5cm}
\begin{equation}
\label{loglFKML}
l_{FKML}(\boldsymbol{\lambda}) = \sum_{i=1}^{n} \log \Bigg[  \frac{\lambda_{2}}{u_{i}^{\lambda_{3} - 1} + (1 - u_{i})^{\lambda_{4}-1}}\Bigg], \lambda_{1},\lambda_{3}, \lambda_{4} \in \mathbb{R}, \lambda_{2} > 0
\vspace{-0.5cm}
\end{equation}
in which $u_{i} = F_{\boldsymbol{\lambda}}(x_{i}), i = 1, \dots, n$. The maximization of (\ref{loglFKML}) is performed applying an algorithm slightly different from the one applied to maximize (\ref{loglRS}). The main issue in maximizing (\ref{loglFKML}) is also in finding initial values for $\{u_{1},\dots,u_{n}\}$. The estimation method, apart from the \mls, that seems to perform best under the \fgld is the method of moments, as outlined by the simulation studies of \citet{lakhany2000}. Therefore, this is the method we use to find the initial values of $\{u_{1},\dots,u_{n}\}$ in a similar manner of what has been done for the \rgld.

The method of moments for the \fglds, as presented in \citet{lakhany2000}, consists on matching the first four sample moments of $\{x_{1},\dots,x_{n}\}$ given by
\vspace{-0.5cm}
\begin{align}
\label{matchFKML}
\hat{\mu}_{1} = & \frac{1}{n} \sum_{i=1}^{n} x_{i} &
\hat{\mu}_{2} = & \frac{1}{n} \sum_{i=1}^{n} (x_{i} - \hat{\mu}_{1})^{2}  \\
\hat{\alpha}_{3} = & \frac{1}{n(\hat{\mu}_{2})^{1.5}} \sum_{i=1}^{n} (x_{i} - \hat{\mu}_{1})^{3} &
\hat{\alpha}_{4} = & \frac{1}{n(\hat{\mu}_{2})^{2}} \sum_{i=1}^{n} (x_{i} - \hat{\mu_{1}})^{4} \nonumber
\vspace{-0.5cm}
\end{align}
to their theoretical moments
\vspace{-0.5cm}
\begin{align}
\label{relationFKML}
\mu_{1}(\boldsymbol{\lambda}) = & \lambda_{1} - \frac{1}{\lambda_{2}}\Big(\frac{1}{\lambda_{3} + 1} - \frac{1}{\lambda_{4} + 1}\Big) &
\mu_{2}(\boldsymbol{\lambda}) = & \frac{1}{\lambda_{2}^{2}}(s_{2} - s_{1}^{2})  \\
\alpha_{3}(\boldsymbol{\lambda}) = & \frac{s_{3} - 3s_{1}s_{2} + 2s_{1}^{3}}{(s_{2} - s_{1}^{2})^{3/2}} &
\alpha_{4}(\boldsymbol{\lambda}) = & \frac{s_{4} - 4s_{1}s_{3} + 6s_{1}^{2}s_{2} - 3s_{1}^{4}}{(s_{2} - s_{1}^{2})^{2}}. \nonumber
\vspace{-0.5cm}
\end{align}
As proposed by \citet{lakhany2000}, we first solve numerically $(\alpha_{3}(\boldsymbol{\lambda}),\alpha_{4}(\boldsymbol{\lambda})) = (\hat{\alpha}_{3}, \hat{\alpha}_{4})$ for $\lambda_{3}$ and $\lambda_{4}$ in the plane $(-1/4,\infty) \times (-1/4,\infty)$ by the minimization of the Euclidean 2-norm $H(\boldsymbol{\lambda}) = \lVert (\alpha_{3}(\boldsymbol{\lambda}),\alpha_{4}(\boldsymbol{\lambda})) - (\hat{\alpha}_{3},\hat{\alpha}_{4}) \rVert_{2}$, and then substitute their values in the first two equations of (\ref{relationFKML}) to obtain $\lambda_{1}$ and $\lambda_{2}$. Using the estimates from the method of moments as initial values, we apply an algorithm analogous to the one applied to the \rgld in order to obtain \ml estimates. The algorithm was also proposed by \citet{su2007}, and is a slight modification of the algorithm of Section 2.1, in which the method of moments is used to find the initial values instead of the percentile method, and the \fgld likelihood is maximized, instead of the \rgld one. More details about it may be found in \citet{su2007}. 

\vskip 4mm

\noindent 3. HURDLE GENERALIZED LAMBDA DISTRIBUTION

In this section we propose a Hurdle Generalized Lambda Distribution (\zglds) for both the RS and \fgld parametrizations, and an estimation technique for its parameters. The \zgld is obtained by adding a fifth parameter $\lambda_{0}$ to either the RS or \fgld that represents their probability mass at zero, so that the hurdle \zgld is a mixed probability distribution.
\vskip 4mm

\noindent 3.1 HURDLE RS GENERALIZED LAMBDA DISTRIBUTION

Let $U$ and $V$ be independent random variables defined in $(\Omega,\mathcal{F},\mathbb{P})$, such that $U$ is uniformly distributed in $[0,1]$ and $\mathbb{P}(V = 1) = 1 - \mathbb{P}(V = 0) = \lambda_{0}$. We say that the random variable $Y_{\boldsymbol{\lambda}^{*}}$ given by
\vspace{-0.5cm}
\begin{equation}
\label{ZIRS}
Y_{\boldsymbol{\lambda}^{*}} \coloneqq Q^{*}_{\boldsymbol{\lambda}^{*}}(U,V) = (1 - V)\Bigg(\lambda_{1} + \frac{U^{\lambda_{3}} - (1 - U)^{\lambda_{4}}}{\lambda_{2}}\Bigg)
\vspace{-0.5cm}
\end{equation}
has a hurdle RS \gld (\zrglds) with parameters $\boldsymbol{\lambda}^{*} = (\lambda_{0}, \lambda_{1}, \lambda_{2}, \lambda_{3}, \lambda_{4})$ in the parametric space $\Lambda^{*} = [0,1] \times \Lambda$.

The random variable $Y_{\boldsymbol{\lambda}^{*}}$ follows a mixed probability distribution, that has a probability mass $\lambda_{0}$ at zero and a probability mass $1 - \lambda_{0}$ spread over $[Q^{*}_{\boldsymbol{\lambda}^{*}}(0,0),Q^{*}_{\boldsymbol{\lambda}^{*}}(1,0)]$ according to an \rglds. As the flexibility of the \gld is maintained in our hurdle generalization, an advantage of fitting an \zrgld is that it is suitable for modelling data with heavy tails and skewness that also has a great quantity of zeros.
 
\vskip 4mm

\noindent 3.1.1 ESTIMATION

The estimation of the \zrgld parameters may be performed by the \ml with an extension of the method of \citet{su2007}. We may represent a sample of $Y_{\boldsymbol{\lambda}^{*}}$ by $\{(y_{1},v_{1}),\dots,(y_{n},v_{n})\}$, in which $y_{i}$ are the observed values and\footnote{$\mathds{1}$ is the indicator function.} $v_{i} = \mathds{1}\{y_{i} = 0\}$, $i=1, \dots,n$, so that the log-likelihood of $\boldsymbol{\lambda}^{*} = (\lambda_{0},\boldsymbol{\lambda})$ is given by
\vspace{-0.5cm}
\begin{equation}
\label{loglZRS}
l_{RS}^{*}(\boldsymbol{\lambda}^{*}) = l_{RS(1)}^{*}(\lambda_{0}) + l_{RS(2)}^{*}(\boldsymbol{\lambda})
\vspace{-0.5cm}
\end{equation}
in which
\vspace{-0.5cm}
\begin{equation*}
\begin{cases}
l_{RS(1)}^{*}(\lambda_{0}) = \sum_{i=1}^{n} v_{i} \log \lambda_{0} + (1-v_{i}) \log (1-\lambda_{0}) \\
l_{RS(2)}^{*}(\boldsymbol{\lambda}) = \sum_{i=1}^{n} (1 - v_{i}) \log \Big[\frac{\lambda_{2}}{\lambda_{3} u_{i}^{\lambda_{3} - 1} + \lambda_{4} (1 - u_{i})^{\lambda_{4}-1}}\Big] 
\end{cases}.
\vspace{-0.1cm}
\end{equation*}
As the log-likelihood (\ref{loglZRS}) may be factored into two functions, one depending on $\lambda_{0}$ and other depending on $\boldsymbol{\lambda}$, the parameters $\lambda_{0}$ and $\boldsymbol{\lambda}$ are orthogonal and, therefore, may be estimated independently. 

The maximum likelihood estimator of $\lambda_{0}$ is $\hat{\lambda_{0}} = \frac{1}{n} \sum_{i=1}^{n} v_{n}$. On the other hand, $(\lambda_{1},\lambda_{2},\lambda_{3},\\\lambda_{4})$ may be estimated by applying the algorithm of Section 2.1 to the non-zero data values, so that we obtain the revised percentile estimator $\hat{\boldsymbol{\lambda}}^{*}$ of the \zrgld under maximum likelihood estimation. As $\hat{\lambda}_{0}$ fits the zero data values perfectly, it is enough to apply diagnostic techniques to the non-zero data values, e.g., by comparing graphically their histogram with the density of an \rgld with parameters $(\hat{\lambda}_{1},\hat{\lambda}_{2},\hat{\lambda}_{3},\hat{\lambda}_{4})$.
\vskip 4mm

\noindent 3.2 HURDLE FKML GENERALIZED LAMBDA DISTRIBUTION

The hurdle FKML \gld (\zfglds) is constructed in the same manner as the \zrglds, by letting $U$ and $V$ be independent random variables defined in $(\Omega,\mathcal{F},\mathbb{P})$, such that $U$ is uniformly distributed in $[0,1]$ and $\mathbb{P}(V = 1) = 1 - \mathbb{P}(V = 0) = \lambda_{0}$, and defining the random variable $Y_{\boldsymbol{\lambda}^{*}}$ as
\vspace{-0.5cm}
\begin{equation}
\label{ZIFKML}
Y_{\boldsymbol{\lambda}^{*}} \coloneqq Q^{*}_{\boldsymbol{\lambda}^{*}}(U,V) = (1 - V)\Bigg( \lambda_{1} + \frac{1}{\lambda_{2}} \Bigg[\frac{U^{\lambda_{3}} - 1}{\lambda_{3}} - \frac{(1 - U)^{\lambda_{4}} - 1}{\lambda_{4}}\Bigg]\Bigg)
\vspace{-0.5cm}
\end{equation}
so that $Y_{\boldsymbol{\lambda}^{*}}$ has an \zfgld with parameters $\boldsymbol{\lambda}^{*} = (\lambda_{0}, \lambda_{1}, \lambda_{2},  \lambda_{3}, \lambda_{4}) \in [0,1] \times \mathbb{R}^{4}$ with the restriction that $\lambda_{2} > 0$ and the same conventions of (\ref{FKML}). The random variable $Y_{\boldsymbol{\lambda}^{*}}$ also follows a mixed probability distribution with the same general characteristics of the \zrglds: it is highly flexible, has a probability mass $\lambda_{0}$ at zero and a probability mass $1 - \lambda_{0}$ spread over $[Q^{*}_{\boldsymbol{\lambda}^{*}}(0,0),Q^{*}_{\boldsymbol{\lambda}^{*}}(1,0)]$ according to an \fglds.
\vskip 4mm

\noindent 3.2.1 ESTIMATION

The estimation of the \zfgld is performed in a way analogous to that of the \zrglds, as the log-likelihood of an \zfgld sample $\{(y_{1},v_{1}), \dots,  (y_{n},v_{n})\}$, $v_{i} = \mathds{1}\{y_{i} = 0\}, i = 1 \dots, n$, may be written as
\vspace{-0.5cm}
\begin{equation}
\label{loglZFKML}
l_{FKML}^{*}(\boldsymbol{\lambda}^{*}) = l_{FKML(1)}^{*}(\lambda_{0}) + l_{FKML(2)}^{*}(\boldsymbol{\lambda}),
\vspace{-0.5cm}
\end{equation}
in which
\vspace{-0.5cm}
\begin{equation*}
\begin{cases}
l_{FKML(1)}^{*}(\lambda_{0}) = \sum_{i=1}^{n} v_{i} \log \lambda_{0} + (1-v_{i}) \log (1-\lambda_{0}) \\
l_{FKML(2)}^{*}(\boldsymbol{\lambda}) = \sum_{i=1}^{n} (1 - v_{i}) \log \Big[\frac{\lambda_{2}}{u_{i}^{\lambda_{3} - 1} + (1 - u_{i})^{\lambda_{4}-1}}\Big],
\end{cases}
\vspace{-0.1cm}
\end{equation*}
so that the parameters $\lambda_{0}$ and $(\lambda_{1},\lambda_{2},\lambda_{3},\lambda_{4})$ are orthogonal, and may be estimated independently.

In order to obtain the revised method of moments estimator $\hat{\boldsymbol{\lambda}}^{*}$ of the \zfgld under maximum likelihood estimation, we estimate $\lambda_{0}$ by the proportion of zero-valued data $\hat{\lambda}_{0} = \frac{1}{n} \sum_{i=1}^{n} v_{i}$ and $(\lambda_{1},\lambda_{2},\lambda_{3},\lambda_{4})$ by the algorithm of Section 2.1, using only the non-zero data values. Diagnostic methods may be applied to the non-zero data values in order to assess the quality of the obtained fit.
\vskip 4mm

\noindent 4. HURDLE GENERALIZED LAMBDA DISTRIBUTION REGRESSION

In this section we propose a regression model for the \zglds, in which we model its location and probability mass at zero as functions of covariates $\boldsymbol{W}$ and $\boldsymbol{Z}$, respectively, which are random vectors defined in $(\Omega,\mathcal{F},\mathbb{P})$, that may share some variables or be equal. Our method is an adaptation of the one presented in \citet{su2015}. We first outline the method of \citet{su2015} and then extend it to the \zglds.
\vskip 4mm

\noindent 4.1 FLEXIBLE PARAMETRIC QUANTILE REGRESSION MODEL

The algorithm of \citet{su2015} seeks to estimate $(\boldsymbol{\beta},\lambda_{2},\lambda_{3},\lambda_{4})$ of the model
\vspace{-0.5cm}
\begin{equation}
\label{reg}
X|\boldsymbol{W}
= \boldsymbol{W}^{T} \boldsymbol{\beta} + \epsilon 
\vspace{-0.5cm}
\end{equation}
in which $\epsilon \sim G\lambda D(\lambda_{1}^{*},\lambda_{2},\lambda_{3},\lambda_{4})$ and $\lambda_{1}^{*}$ is such that $E(\epsilon) = 0$, i.e.,
\vspace{-0.5cm}
\begin{equation}
\label{lstar}
\lambda_{1}^{*} = \begin{cases}
- \frac{(\lambda_{3} + 1)^{-1} - (\lambda_{4} + 1)^{-1}}{\lambda_{2}} & \text{ for the \rglds} \\
\frac{(\lambda_{3} + 1)^{-1} - (\lambda_{4} + 1)^{-1}}{\lambda_{2}} & \text{ for the \fglds}
\end{cases}.
\vspace{-0.5cm}
\end{equation}
In order to estimate the parameters of (\ref{reg}) we apply a 5-step algorithm that is analogous to the algorithms of Section 2.1: find initial values to the parameters in order to evaluate and maximize the log-likelihood to get \ml estimates. It is supposed that we have a sample $\{(x_{1},\boldsymbol{w}_{1}),\dots,(x_{n},\boldsymbol{w}_{n})\}$ of the response variable and covariates. The algorithm is as follows and more details about it are presented in \citet{su2015}.

\begin{enumerate}[nolistsep]
	\item Obtain $\hat{\boldsymbol{\beta}}^{(0)}$ from the least square method by solving
	\vspace{-0.5cm}
	\begin{equation*}
	\hat{\boldsymbol{\beta}}^{(0)} = \arg\min\limits_{\boldsymbol{\beta}} \sum_{i=1}^{n} \big(x_{i} - \boldsymbol{w_{i}}^{T} \boldsymbol{\beta}\big)^{2}
	\vspace{-0.5cm}
	\end{equation*}
	and calculate the initial residuals $\hat{e}_{i}^{(0)} = x_{i} - \boldsymbol{w_{i}}^{T} \hat{\boldsymbol{\beta}}^{(0)}$.
	\item Obtain the initial estimates $(\hat{\lambda}_{2}^{(0)},\hat{\lambda}_{3}^{(0)},\hat{\lambda}_{4}^{(0)})$ by applying the algorithm of Section 2.1 to sample $\{e_{1}^{(0)},\dots,e_{n}^{(0)}\}$ of $\epsilon$.
	\item Calculate the log-likelihood of the model as follows:
	\begin{itemize}
		\item[(a)] Evaluate $\hat{\lambda}_{1}^{*(0)}$ by (\ref{lstar}) so that the initial estimated distribution of the error $\epsilon$ has zero mean.
		\item[(b)] Force the residuals sample mean to be zero by making
		\vspace{-0.5cm}
		\begin{equation*}
		e_{i}^{*} = (y_{i} - \boldsymbol{w}_{i}^{T}\boldsymbol{\beta}) - \frac{1}{n} \sum_{i=1}^{n} e_{i}^{*}
		\vspace{-0.5cm}
		\end{equation*}
		\item[(c)] Evaluate the log-likelihood of the zero mean residuals from equations (\ref{loglRS}) or (\ref{loglFKML}):
		\begin{itemize}
			\item[(i)] For the \rgld with $\boldsymbol{\lambda} \in \Lambda$ 
			\vspace{-0.5cm}
			\begin{align}
			\label{llRS}
			& l_{e^{*}}(\boldsymbol{\beta},\lambda_{2},\lambda_{3},\lambda_{4}) = \sum_{i=1}^{n} \log \Bigg[  \frac{\lambda_{2}}{\lambda_{3} u_{i}^{\lambda_{3} - 1} + \lambda_{4} (1 - u_{i})^{\lambda_{4}-1}}\Bigg] \\
			\label{l1}
			& e_{i}^{*} = \lambda_{1}^{*} + \frac{u_{i}^{\lambda_{3}} - (1 - u_{i})^{\lambda_{4}}}{\lambda_{2}}
			\vspace{-0.5cm}
			\end{align}
			\item[(ii)] For the \fgld with $\lambda_{3}, \lambda_{4} \in \mathbb{R}, \lambda_{2} > 0$
			\vspace{-0.5cm}
			\begin{align}
			\label{llFKML}
			& l_{e^{*}}(\boldsymbol{\beta},\lambda_{2},\lambda_{3},\lambda_{4}) = \sum_{i=1}^{n} \log \Bigg[  \frac{\lambda_{2}}{u_{i}^{\lambda_{3} - 1} + (1 - u_{i})^{\lambda_{4}-1}}\Bigg] \\
			\label{l2}
			& e_{i}^{*} = \lambda_{1}^{*} + \frac{1}{\lambda_{2}} \Bigg[\frac{u_{i}^{\lambda_{3}} - 1}{\lambda_{3}} - \frac{(1 - u_{i})^{\lambda_{4}} - 1}{\lambda_{4}}\Bigg]
			\vspace{-0.5cm}
			\end{align}
		\end{itemize}
		in which $u_{i}$ is given implicitly by (\ref{l1}) and (\ref{l2}), depending on the parametrization, and is a function of $(\boldsymbol{\beta},\lambda_{2},\lambda_{3},\lambda_{4})$.
	\end{itemize}
	\item[4.] Maximize numerically, by the Nelder-Mead simplex algorithm \citep{nelder1965}, for example, the log-likelihood (\ref{llRS}) or (\ref{llFKML}), depending on the parametrization, using $\hat{\boldsymbol{\beta}}^{(0)}, \hat{\lambda}_{2}^{(0)}, \hat{\lambda}_{3}^{(0)}$ and $\hat{\lambda}_{4}^{(0)}$ as initial values, in order to obtain $\hat{\boldsymbol{\beta}}, \hat{\lambda}_{2}, \hat{\lambda_{3}}$ and $\hat{\lambda}_{4}$. 
	\item[5.] Obtain $\hat{\lambda}_{1}^{*}$ substituting the estimated values $\hat{\lambda}_{2}, \hat{\lambda_{3}}$ and $\hat{\lambda}_{4}$ in (\ref{lstar}).
	\item[6.] Conduct simulations to obtain statistical properties of the estimated regression coefficients $\hat{\beta}$ as follows:
	\begin{itemize}
		\item[(a)] Generate $\{\epsilon_{1},\dots,\epsilon_{n}\}$ from the \gld with parameters $(\hat{\lambda}_{1}^{*},\hat{\lambda}_{2},\hat{\lambda}_{3},\hat{\lambda}_{4})$ and obtain a new sample $\{y_{1}^{*},\dots,y_{n}^{*}\}$ by adding $y_{i}^{*} = \boldsymbol{w_{i}}^{T} \hat{\boldsymbol{\beta}} + \epsilon_{i}, i = 1,\dots,n$. Fit a regression model to $\{y^{*}_{1},\dots,y^{*}_{n}\}$ obtaining estimates for the regression coefficients.
		\item[(b)] Repeat step (a) $1,000$ times to obtain $1,000$ coefficients\footnote{The number $1,000$ is arbitrary. It could be sampled more or less coefficients.}.
		\item[(c)] Adjust the each coefficient sample in (b) so that its mean is equal to the final estimated coefficients of step 5. The simulated coefficients histogram may be plotted and $(1-\alpha)\%$ confidence intervals may be found by evaluating the $\alpha/2$ and $1-\alpha/2$ quantiles of the simulated samples. We use quantile type 8 from the quantile function in \textbf{R} \citep{hyndman1996,R} in order to be consistent to \citet{su2015}.
	\end{itemize}
\end{enumerate}

Any other method could be used to estimate the parameters of the error distribution in step 2. However, we prefer the \ml for it provides better estimates, as has been established on the literature, although it may not converge in some cases. A limitation of this method is the lack of asymptotic theoretical results about the distribution of the estimators, so that we cannot construct asymptotic confidence intervals, nor test hypothesis, for the coefficients. Nevertheless, computational methods for generating confidence intervals for the coefficients and for establishing goodness of fit are implemented and can be applied \citep{su2016}.

\vskip 4mm

\noindent 4.2 \zgld REGRESSION MODEL

In order to develop an \zgld regression model, we rely on the factorization of the log-likelihoods (\ref{loglZRS}) and (\ref{loglZFKML}), as it allows to model the parameter $\lambda_{0}$ and the location of the distribution independently. Indeed, our regression model, whose response variable is $Y$ and covariates are\footnote{Note that $\boldsymbol{W}$ and $\boldsymbol{Z}$ may share some of the same variables or be equal.} $(\boldsymbol{W},\boldsymbol{Z})$, may be written as
\vspace{-0.5cm}
\begin{equation}
\label{modelZI}
\begin{cases}
Y|(\boldsymbol{W},\boldsymbol{Z}) = (1 - (V|\boldsymbol{Z}))(\boldsymbol{W}^{T} \boldsymbol{\beta} + \epsilon) \\
\log \bigg(\frac{\mathbb{P}(V = 1|\boldsymbol{Z})}{1 - \mathbb{P}(V = 1|\boldsymbol{Z})}\bigg) = \boldsymbol{Z}^{T} \boldsymbol{\gamma}
\end{cases}
\vspace{-0.5cm}
\end{equation}
in which $\epsilon \sim G\lambda D(\lambda_{1}^{*},\lambda_{2},\lambda_{3},\lambda_{4})$ and $\lambda_{1}^{*}$ is such that $E(\epsilon) = 0$, i.e., is given by relation (\ref{lstar}). 

Given a sample $\{(y_{1},v_{1},\boldsymbol{w}_{1},\boldsymbol{z}_{1}), \dots, (y_{n},v_{n},\boldsymbol{w}_{n},\boldsymbol{z}_{n})\}$ of model (\ref{modelZI}), in which $v_{i} = \mathds{1}\{y_{i} = 0\}$, the log-likelihood of the parameters is given by
\vspace{-0.5cm}
\begin{align}
\label{llZIR} \nonumber
& l(\boldsymbol{\beta},\boldsymbol{\gamma},\lambda_{2},\lambda_{3},\lambda_{4}) = \sum_{i=1}^{n} \log \Bigg( \Bigg[\frac{\exp(\boldsymbol{z}_{i}^{T} \boldsymbol{\gamma})}{1+\exp(\boldsymbol{z}_{i}^{T} \boldsymbol{\gamma})}\Bigg]^{v_{i}} \Bigg[\frac{f(y_{i} - \boldsymbol{w}_{i}^{T} \boldsymbol{\beta})}{1 + \exp(\boldsymbol{z}_{i}^{T} \boldsymbol{\gamma})}\Bigg]^{(1-v_{i})} \Bigg) \\ 
& = \sum_{i=1}^{n} v_{i}\boldsymbol{z}_{i}^{T} \boldsymbol{\gamma} - \log(1 + \exp(\boldsymbol{z}_{i}^{T} \boldsymbol{\gamma})) + (1-v_{i})\log(f_{(\lambda_{1}^{*},\lambda_{2},\lambda_{3},\lambda_{4})}(y_{i} - \boldsymbol{w}_{i}^{T}\boldsymbol{\beta})) \nonumber \\
& \coloneqq l_{1}^{*}(\boldsymbol{\gamma}) + l_{2}^{*}(\boldsymbol{\beta},\lambda_{1}^{*},\lambda_{2},\lambda_{3},\lambda_{4})
\vspace{-0.5cm}
\end{align}
in which $f(y_{i} - \boldsymbol{w}_{i}^{T}\boldsymbol{\beta})$ is either the density (\ref{RSdensity}) or (\ref{FKMLdensity}) with parameters $(\lambda_{1}^{*},\lambda_{2},\lambda_{3},\lambda_{4})$ evaluated at point $y_{i} - \boldsymbol{w}_{i}^{T}\boldsymbol{\beta}, i = 1,\dots,n$.

The estimation of the parameters of model (\ref{modelZI}) may be performed by maximizing $l_{1}^{*}(\boldsymbol{\gamma})$ and $l_{2}^{*}(\boldsymbol{\beta},\lambda_{1}^{*},\lambda_{2},\lambda_{3},\lambda_{4})$ independently, so that we get the maximum likelihood estimator $\hat{\boldsymbol{\gamma}}$ and the \ml estimators $\hat{\boldsymbol{\beta}}, \hat{\lambda}_{1}^{*}, \hat{\lambda_{2}}, \hat{\lambda}_{3}$ and $\hat{\lambda}_{4}$. On the one hand, the maximization of $l_{1}^{*}(\boldsymbol{\gamma})$ is performed by fitting a logistic regression in the usual manner, as shown in \citet{hilbe2009} for example, to sample $\{(v_{1},\boldsymbol{z}_{1}), \dots, (v_{n},\boldsymbol{z}_{n})\}$. On the other hand, the maximization of $l_{2}^{*}(\boldsymbol{\beta}, \lambda_{1}^{*}, \lambda_{2}, \lambda_{3}, \lambda_{4})$ is performed by applying the algorithm of Section 4.1 to the non-zero data values.

As the parameters $\boldsymbol{\gamma}$ and $(\boldsymbol{\beta},\lambda_{1}^{*},\lambda_{2},\lambda_{3},\lambda_{4})$ are orthogonal, their maximum likelihood estimators are asymptotically independent \citep{cox1987}. Therefore, the usual methods of inference for logistic regression models may be applied to infer about $\boldsymbol{\gamma}$. Similarly, logistic regression diagnostic techniques may also be applied in order to asses the quality of the fit. However, as the estimators $\hat{\boldsymbol{\beta}}, \hat{\lambda}_{1}^{*}, \hat{\lambda_{2}}, \hat{\lambda}_{3}$ and $\hat{\lambda}_{4}$ are not of maximum likelihood, the usual inference techniques for maximum likelihood estimators cannot be applied to them. Nevertheless, we may construct numerical confidence intervals for $\boldsymbol{\beta}$ by applying the method of step 6 of algorithm of Section 4.1 to the non-zero data values. 

The goodness-of-fit of \zgld regression models may be established by the study of two kinds of residuals: error residuals and normalized quantile residuals. The error residuals are given by $e = y - \boldsymbol{w}^{T}\boldsymbol{\hat{\beta}}$ for all $y \neq 0$ and their empirical distribution may be compared with the \glds$(\hat{\lambda}_{1}^{*}, \hat{\lambda_{2}}, \hat{\lambda}_{3}, \hat{\lambda}_{4})$, that was fitted to the error $\epsilon$ in order to establish goodness-of-fit. This comparison may be performed by the use of QQ-plots, a histogram of $e$ superimposed by the estimated density and a quantile plot that superimposes the estimated and the empirical quantile functions of $\epsilon$ and $e$, respectively.

The normalized quantile residuals, as presented, for example, in \citet{quantileresiduals}, are defined as $r = \Phi^{-1}(F_{\boldsymbol{\lambda}}^{*}(y - \boldsymbol{w}^{T}\boldsymbol{\hat{\beta}}))$, in which $\Phi$ and $F_{\boldsymbol{\lambda}}^{*}$ are the cumulative distribution function of the standard normal distribution and the RS or \fglds$(\hat{\lambda}_{1}^{*}, \hat{\lambda_{2}}, \hat{\lambda}_{3}, \hat{\lambda}_{4})$, respectively. The normalized quantile residuals are expected to be normally distributed if the model is properly fitted, so that we may regard the model as well fitted if the density estimate of $r$ is close to the standard normal distribution density and the points of the normal QQ-plot of $r$ are distributed around the line with intercept zero and slope one, for example. These residuals may also be used to asses the goodness-of-fit of the logistic regression model \citep{gamlss}.
\vskip 4mm

\noindent 5. SIMULATION STUDY

In this section we perform a simulation study in order to assess the asymptotic properties of the \zgld regression models. We consider the model
\vspace{-0.5cm}
\begin{equation}
\label{model_simulation}
\begin{cases}
Y|(x_{1},x_{2}) = (1 - (V|x_{1},x_{2}))(6.13 - 0.021x_{1} - 0.35x_{2} + \epsilon) \\
\log \bigg(\frac{\mathbb{P}(V = 1|x_{1},x_{2})}{1 - \mathbb{P}(V = 1|x_{1},x_{2})}\bigg) = 1.6 - 0.13x_{1} + 0.21x_{2}
\end{cases}
\vspace{-0.5cm}
\end{equation}
in which $x_{1} \sim RS \ G\lambda D(3.87,0.10,0.024,0.19)$ and $\mathbb{P}(x_{2} = 1) = 1 - \mathbb{P}(x_{2} = 0) = 0.6$. We consider four different scenarios in our simulations, in which the distribution of the $\epsilon$ error is symmetric (\rglds(0,2,0.13,0.13) and \fglds(0,2,0.13,0.13)), and right skewed (\rglds(-1.43,0.11,0.0023,0.19) and \fglds(-0.147,-0.41,1.07,0.84,0.02)). In each scenario, we generate $1,000$ samples of model (\ref{model_simulation}), for each sample size $n = 100, 200$ and $1,000$, and, for each sample, we fit a \zgld regression model, estimating the coefficients of (\ref{model_simulation}). We them study the mean, standard error and the 2.5th and 97.5th percentiles of the estimated coefficients of (\ref{model_simulation}) over $1,000$ samples. The results are presented in Table \ref{simulation}.

We observe that, in all scenarios, the mean of the estimated coefficients is close to the target value, especially for the sample of size $n = 1,000$, which is evidence that the estimators are unbiased. Furthermore, we see that as greater the sample size is, smaller is the standard error of the estimated coefficients, which is evidence that the estimators are consistent. Overall, the simulation study support the consistency of the estimators, so that it is not lost when we consider the hurdle model: the logistic regression consistency, theoretically established, and the consistency of the \gld regression, supported by the simulations of \cite{su2015}, seems to be preserved when we consider the hurdle model.

\begin{table}
	\centering
	\caption{Mean, standard error and the 2.5th and 97.5th percentiles of samples of the coefficients of (\ref{model_simulation}).}\label{simulation}
	\resizebox{\linewidth}{0.49\textheight}{
	\begin{tabular}{cccccccc}
	\hline
	\multirow{2}{*}{Distribution of $\epsilon$} &\multirow{2}{*}{Coefficient} & \multirow{2}{*}{Target} & Sample & \multirow{2}{*}{Mean} & Standard & \multicolumn{2}{c}{Percentiles} \\ \cline{7-8}
	& & & Size & & Error & 2.5th & 97.5th \\  
	\hline
	& \multirow{3}{*}{Non-zero intercept} & \multirow{3}{*}{6.12} & 100 & 6.131 & 0.091 & 5.959 & 6.309 \\ 
	&  & & 200 & 6.130 & 0.059 & 6.009 & 6.244 \\ 
	& & & 1,000 & 6.130 & 0.022 & 6.085 & 6.174 \\ \cline{2-8}
	& \multirow{3}{*}{Non-zero $x_{1}$} & \multirow{3}{*}{-0.021} & 100 & -0.021 & 0.016 & -0.052 & 0.008 \\ 
	&  & & 200 & -0.021 & 0.010 & -0.040 & 0.001 \\
	& & & 1,000 & -0.021 & 0.004 & -0.028 & -0.013 \\ \cline{2-8}
	& \multirow{3}{*}{Non-zero $x_{2}$} & \multirow{3}{*}{-0.35} & 100 & -0.350 & 0.048 & -0.446 & -0.255 \\ 
	&  & & 200 & -0.350 & 0.033 & -0.416 & -0.284 \\
	HRS& & & 1,000 & -0.350 & 0.012 & -0.375 & -0.326 \\ \cline{2-8}
	\glds(0,2,0.13,0.13)& \multirow{3}{*}{Zero intercept} & \multirow{3}{*}{1.6} & 100 & 1.659 & 0.944 & -0.001 & 3.762 \\ 
	&  & & 200 & 1.627 & 0.650 & 0.401 & 2.984 \\ 
	& & & 1,000 & 1.611 & 0.272 & 1.097 & 2.136 \\ \cline{2-8}
	& \multirow{3}{*}{Zero $x_{1}$} & \multirow{3}{*}{-0.13} & 100 & -0.136 & 0.164 & -0.464 & 0.191 \\ 
	&  & & 200 & -0.131 & 0.110 & -0.362 & 0.083 \\
	& & & 1,000 & -0.132 & 0.046 & -0.223 & -0.043 \\  \cline{2-8}
	& \multirow{3}{*}{Zero $x_{2}$} & \multirow{3}{*}{0.21} & 100 & 0.190 & 0.492 & -0.768 & 1.139 \\ 
	&  & & 200 & 0.207 & 0.341 & -0.481 & 0.829 \\
	& & & 1,000 & 0.216 & 0.149 & -0.072 & 0.509 \\ 
	\hline	 	
	& \multirow{3}{*}{Non-zero intercept} & \multirow{3}{*}{6.12} & 100 & 6.145 & 0.750 & 4.615 & 7.664 \\ 
	&  & & 200 & 6.114 & 0.489 & 5.215 & 7.215 \\ 
	& & & 1,000 & 6.123 & 0.176 & 5.780 & 6.470 \\ \cline{2-8}
	& \multirow{3}{*}{Non-zero $x_{1}$} & \multirow{3}{*}{-0.021} & 100 & -0.022 & 0.129 & -0.283 & 0.243 \\ 
	& & & 200 & -0.018 & 0.083 & -0.201 & 0.131 \\ 
	& & & 1,000 & -0.020 & 0.029 & -0.077 & 0.034 \\   \cline{2-8}
	& \multirow{3}{*}{Non-zero $x_{2}$} & \multirow{3}{*}{-0.35} & 100 & -0.359 & 0.409 & -1.124 & 0.469 \\ 
	&  & & 200 & -0.353 & 0.266 & -0.876 & 0.165 \\ 
	HFKML& & & 1,000 & -0.346 & 0.094 & -0.532 & -0.158 \\ \cline{2-8} 	
	\glds(0,2,0.13,0.13)& \multirow{3}{*}{Zero intercept} & \multirow{3}{*}{1.6} & 100 & 1.623 & 0.916 & 0.040 & 3.577 \\ 
	&  & & 200 & 1.620 & 0.649 & 0.388 & 2.975 \\ 
	& & & 1,000 & 1.611 & 0.272 & 1.097 & 2.136 \\  \cline{2-8}
	& \multirow{3}{*}{Zero $x_{1}$} & \multirow{3}{*}{-0.13} & 100 & -0.136 & 0.159 & -0.438 & 0.171 \\ 
	& & & 200 & -0.131 & 0.108 & -0.362 & 0.081 \\
	& & & 1,000 & -0.132 & 0.046 & -0.223 & -0.043 \\  \cline{2-8}
	& \multirow{3}{*}{Zero $x_{2}$} & \multirow{3}{*}{0.21} & 100 & 0.206 & 0.482 & -0.735 & 1.149 \\ 
	&  & & 200 & 0.205 & 0.342 & -0.484 & 0.834 \\ 
	& & & 1,000 & 0.216 & 0.149 & -0.072 & 0.509 \\ 
	\hline
	& \multirow{3}{*}{Non-zero intercept} & \multirow{3}{*}{6.12} & 100 & 6.080 & 0.784 & 4.434 & 7.697 \\ 
	& & & 200 & 6.102 & 0.448 & 5.200 & 7.003 \\
	& & & 1,000 & 6.115 & 0.179 & 5.742 & 6.451 \\ \cline{2-8}
	& \multirow{3}{*}{Non-zero $x_{1}$} & \multirow{3}{*}{-0.021} & 100 & -0.014 & 0.134 & -0.273 & 0.300 \\ 
	& & & 200 & -0.019 & 0.073 & -0.158 & 0.134 \\
	& & & 1,000 & -0.020 & 0.028 & -0.080 & 0.041 \\  \cline{2-8}
	& \multirow{3}{*}{Non-zero $x_{2}$} & \multirow{3}{*}{-0.35} & 100 & -0.337 & 0.403 & -1.197 & 0.518 \\ 
	&  & & 200 & -0.339 & 0.210 & -0.738 & 0.078 \\
	HRS& & & 1,000 & -0.342 & 0.067 & -0.462 & -0.185 \\  \cline{2-8}	
	\glds(-1.43,0.11,0.0023,0.19)& \multirow{3}{*}{Zero intercept} & \multirow{3}{*}{1.6} & 100 & 1.623 & 0.906 & -0.045 & 3.534 \\ 
	& & & 200 & 1.632 & 0.620 & 0.433 & 2.924 \\ 
	& & & 1,000 & 1.616 & 0.273 & 1.068 & 2.152 \\ \cline{2-8}
	& \multirow{3}{*}{Zero $x_{1}$} & \multirow{3}{*}{-0.13} & 100 & -0.131 & 0.160 & -0.442 & 0.188 \\ 
	& & & 200 & -0.134 & 0.107 & -0.349 & 0.083 \\ 
	& & & 1,000 & -0.132 & 0.047 & -0.223 & -0.038 \\   \cline{2-8}
	& \multirow{3}{*}{Zero $x_{2}$} & \multirow{3}{*}{0.21} & 100 & 0.211 & 0.493 & -0.776 & 1.171 \\ 
	& & & 200 & 0.209 & 0.328 & -0.447 & 0.848 \\ 
	& & & 1,000 & 0.213 & 0.155 & -0.073 & 0.507 \\
	\hline	 	
	& \multirow{3}{*}{Non-zero intercept} & \multirow{3}{*}{6.12} & 100 & 6.139 & 0.777 & 4.584 & 7.906 \\ 
	& & & 200 & 6.083 & 0.411 & 5.259 & 6.861 \\ 
	& & & 1,000 & 6.130 & 0.110 & 5.923 & 6.350 \\ \cline{2-8}
	& \multirow{3}{*}{Non-zero $x_{1}$} & \multirow{3}{*}{-0.021} & 100 & -0.022 & 0.133 & -0.309 & 0.236 \\ 
	& & & 200 & -0.011 & 0.066 & -0.149 & 0.128 \\
	& & & 1,000 & -0.021 & 0.015 & -0.050 & 0.011 \\  \cline{2-8}%%
	& \multirow{3}{*}{Non-zero $x_{2}$} & \multirow{3}{*}{-0.35} & 100 & -0.352 & 0.408 & -1.232 & 0.513 \\ 
	& & & 200 & -0.366 & 0.187 & -0.742 & 0.019 \\
	HFKML& & & 1,000 & -0.351 & 0.042 & -0.446 & -0.268 \\  \cline{2-8}	
	\glds(-0.147,-0.41,1.07,0.84,0.02)& \multirow{3}{*}{Zero intercept} & \multirow{3}{*}{1.6} & 100 & 1.635 & 0.916 & -0.098 & 3.433 \\  
	&  & & 200 & 1.615 & 0.652 & 0.402 & 2.962 \\
	& & & 1,000 & 1.613 & 0.279 & 1.054 & 2.154 \\ \cline{2-8}
	& \multirow{3}{*}{Zero $x_{1}$} & \multirow{3}{*}{-0.13} & 100 & -0.132 & 0.157 & -0.434 & 0.187 \\ 
	& & & 200 & -0.132 & 0.111 & -0.353 & 0.086 \\  
	& & & 1,000 & -0.132 & 0.048 & -0.225 & -0.035 \\ \cline{2-8}
	& \multirow{3}{*}{Zero $x_{2}$} & \multirow{3}{*}{0.21} & 100 & 0.194 & 0.509 & -0.820 & 1.246 \\ 
	& & & 200 & 0.200 & 0.338 & -0.459 & 0.847 \\
	& & & 1,000 & 0.212 & 0.154 & -0.073 & 0.507 \\ 
	\hline
	\end{tabular}}
\end{table}

%\vskip 4mm
\newpage
\noindent 6. FITTING AN \zgld TO HEALTHCARE EXPENSES DATA

Healthcare expenses data has some peculiarities which make the \zgld a great option for modelling it. Indeed, yearly healthcare expenses data has usually a great number of zeros, normally more than 50\% of the data, as not every person uses their health insurance in the period of a year. Furthermore, the distribution of healthcare expenses is highly skewed and has a heavy tail that is hardly modelled by the usual distributions, as the Gamma, Weibull, Log-normal and Inverse-Gaussian.

In the following sections, we fit models to a dataset that contains the yearly expenses of all insured customers of a Brazilian healthcare insurance company between 2006 and 2009. Our analysis focuses on modelling the yearly expenses in function of the covariates age, sex and previous year expenses. All expenses are in \textit{Reais}\footnote{Brazilian currency.} (R\$) and were deflated to January 2006 value. The \zgld models are compared with \gpd models in order to establish which model best fits the data.

The \gpds, introduced by \citet{pickands1975}, is a three parameter positive probability distribution with density
\vspace{-0.5cm}
\begin{equation*}
f(y) = \begin{cases}
\frac{1}{\tau} \big(1 + \xi \frac{y - \alpha}{\tau}\big)^{-\frac{\xi + 1}{\xi}} & \xi \neq 0 \\
\frac{1}{\tau} \exp\big(-\frac{y - \alpha}{\tau}\big) & \xi = 0
\end{cases}
\vspace{-0.3cm}
\end{equation*}
for $y \geq \alpha$, in which $\alpha \geq 0$ is the location parameter, $\tau > 0$ is the scale parameter and $\xi \in \mathbb{R}$ is the shape parameter. The mean of the \gpd is finite only for $\xi < 1$ and is given by
\vspace{-0.5cm}
\begin{equation}
\label{meanGPD}
E(Y) \coloneqq \mu = \alpha + \frac{\tau}{1 - \xi}.
\vspace{-0.5cm}
\end{equation}
Note that the \gpd may be re-parametrized so that $\mu$ is the scale parameter, instead of $\tau$. Given a sample $\{y_{1},\dots,y_{n}\}$ and a known threshold $\alpha$, the parameters $(\xi,\tau)$ (or $(\xi,\mu)$) may be estimated by the Maximum Likelihood Method in the usual manner. See \citet{hosking1987} and \citet{grimshaw1993} for more details.

In order to fit a \gpd to data when there are covariates, we may use a generalized linear model (GLM) framework, as introduced by \citet{nelder1972}. In this framework, we suppose that the location parameter $\alpha$ is known and independent of the covariates, and that the shape parameter $\xi$ is unknown, but is lesser than one and independent of the covariates. Then, we model the mean $\mu$ as $E(Y|\boldsymbol{x}_{i}) \coloneqq \mu_{i} = \exp(\boldsymbol{x}_{i}^{T} \boldsymbol{\beta})$, in which $\boldsymbol{x}_{i}$ are the covariates of the $i-th$ observation and $\boldsymbol{\beta}$ are the coefficients of the model. The coefficients $(\xi,\boldsymbol{\beta})$ are estimated by Maximum Likelihood numerically and their asymptotic distributions are obtained by the asymptotic properties of Maximum Likelihood Estimators. 

A Hurdle Generalized Pareto Distribution (HGPD) model may be developed in a similar manner as the \zgld model. Indeed, it is enough to add a parameter $\lambda_{0}$ to the \gpd that represents its probability mass at zero and then estimate the parameters accordingly: the estimate of $\lambda_{0}$ is the proportion of zeros in the sample and the estimate of $(\xi,\tau)$ is the Maximum Likelihood estimate for the \gpd fitted to the non-zero data values. A HGPD GLM is obtained by replacing $(\boldsymbol{W}^{T}\boldsymbol{\beta} + \epsilon)$ in expression (\ref{modelZI}) by a random variable $U|\boldsymbol{W}^{T}$ that has a \gpd with parameters $(\alpha,\xi,\mu = \exp(\boldsymbol{W}^{T}\boldsymbol{\beta}))$. The parameters related to the probability mass at zero and to the \gpd for the non-zero values are orthogonal, so that their estimation may be performed independently, as was the case of the \zglds. This hurdle model is a special case of the Zero-inflated Truncated Generalized Pareto Distribution model introduced by \citet{couturier2010}.

In order to establish the goodness-of-fit of the HGPD GLM we may consider the zero and non-zero data values separately. For the zero values we consider logistic regression diagnostic techniques and for the non-zero values we propose the study of two types of residuals: normalised quantile residuals and error residuals, that are given respectively by
\vspace{-0.5cm}
\begin{align*}
r = \Phi^{-1}(F_{(\alpha,\hat{\xi},\hat{\mu})}(y)) & & e = \frac{y - \alpha}{\hat{\mu}}
\vspace{-0.5cm}
\end{align*}
in which $F_{(\alpha,\hat{\xi},\hat{\mu})}$ is the cumulative probability function of a GPD with parameters $(\alpha,\hat{\xi},\hat{\mu})$, for $y \neq 0$. If the model is well-fitted then $r$ is normally distributed and $e$ has a \gpd with parameters $\alpha_{e} = 0$, $\xi_{e} = \hat{\xi}$ and $\mu_{e} = 1$, so that graphical tools, as QQ-plots, may be used to establish goodness-of-fit. The error residuals were proposed by \citet{couturier2010} where more details are presented.

\vskip 4mm
\FloatBarrier
\noindent 6.1 THE DATASET

\begin{figure}[t]
	\centering
	\includegraphics[width = 0.75\linewidth, height=0.2\textheight]{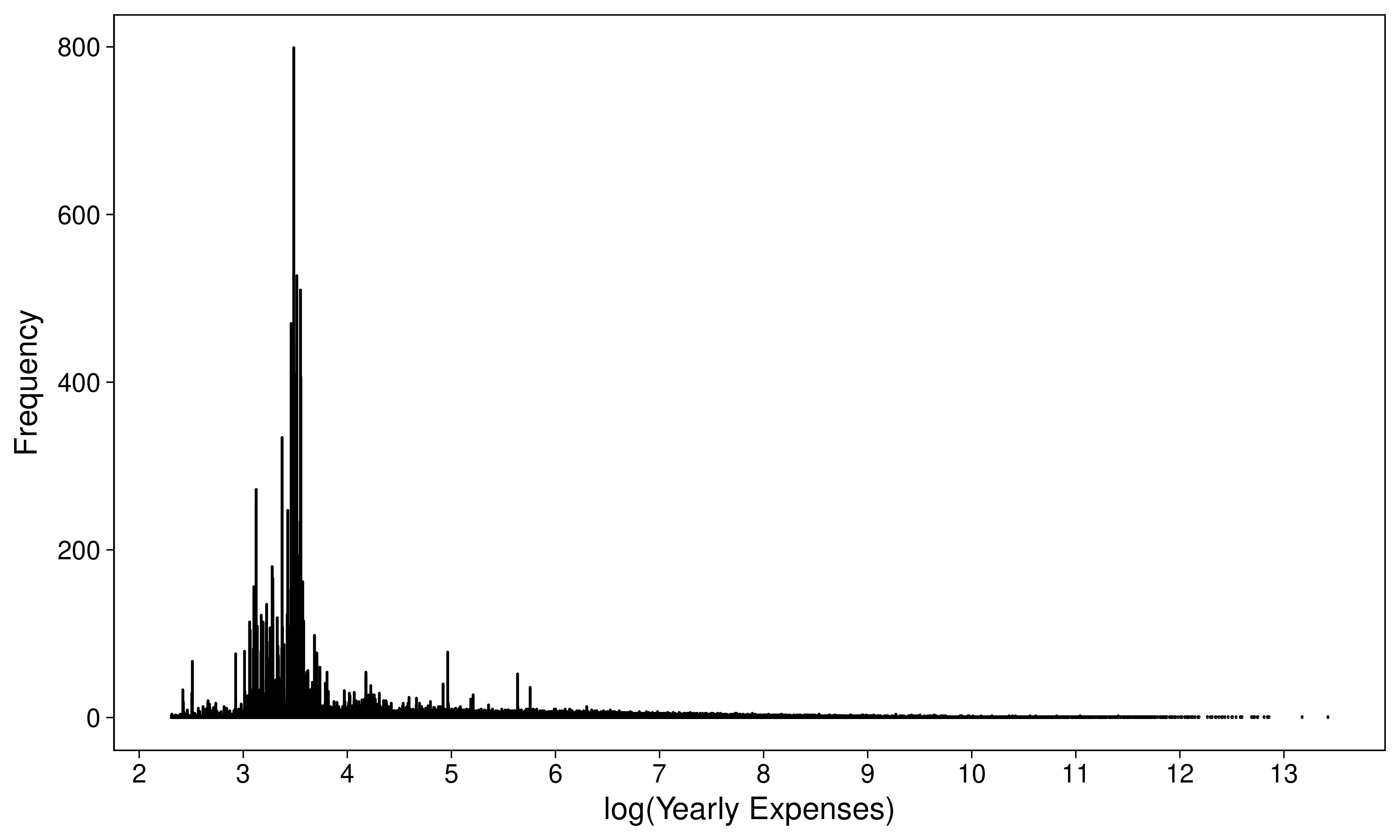}
	\caption{Frequency of each yearly expense value greater than zero in the logarithm scale.}
	\label{hist}
\end{figure}

In order to fit a model to the data at hand, we first observe some systematic behaviour of the data and transform it to obtain a better fit. First of all, there are some yearly expense values that are observed in the dataset hundreds of times, as can be seen in Figure \ref{hist}, as there are some simple medical procedures that have standardized costs. Those repeated values make it hard to fit a continuous model, as some values have a probability mass greater than zero. Therefore, we consider that any expense less than R\$ 100 is zero, i.e., we truncate the yearly expenses at R\$ 100, and consider all yearly expenses lesser than R\$ 100 to be zero. This truncation is justified by the practical application of the fitted model, as the main interest in modelling healthcare expenses is in properly fitting the tail of the distribution, i.e., the yearly expenses that are dozens of times the expected one, so that low expenses, as those less than R\$ 100, may be regarded as zero without any loss for the practical application of the model. Indeed, around 69 \% of the dataset has an expense less than R\$ 100, although their expenses sum to R\$ 2,552,800, that is less then 2\% of the total expenses of the dataset, that is R\$ 137,382,575.

\begin{figure}[t]
	\centering
	\includegraphics[width = 0.75\linewidth, height=0.2\textheight]{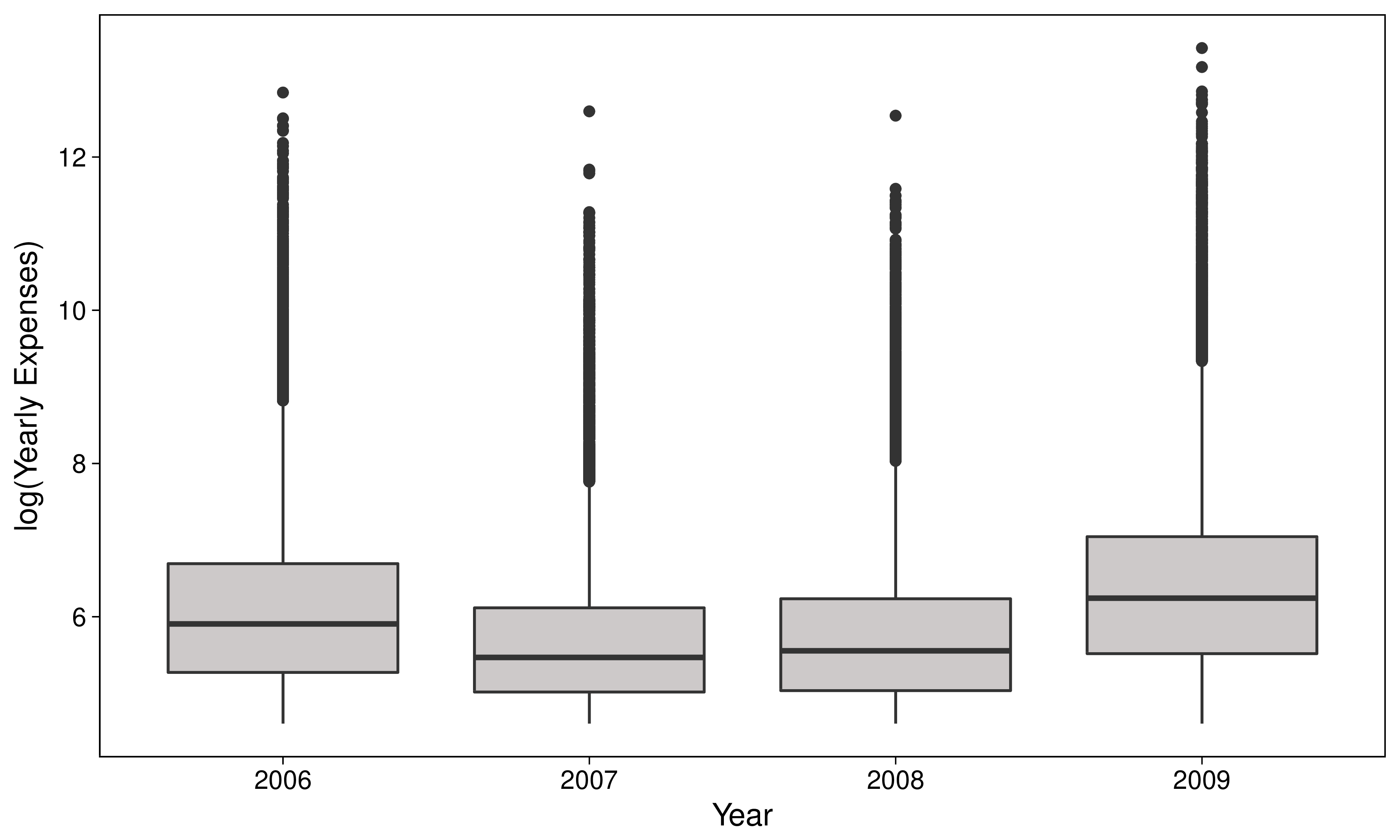}
	\caption{Box-plot of the logarithm of the yearly expenses by year. The expenses lesser than R\$ 100 were omitted for a better visualization.}
	\label{box-plot}
\end{figure}

Truncating the dataset at R\$ 100, we have, for each year and for the whole dataset, the proportion of zeros, selected percentiles, mean, standard deviation and maximum expense displayed in Table \ref{descriptive}. The percentiles, mean and standard deviation refer to the truncated data, i.e., are calculated considering only data values greater than R\$ 100. From Table \ref{descriptive} it can be seen that the 99th percentile is approximately twice the 98th percentile, the same occurring with the 99th and 99.5th percentiles. Furthermore, the 99.9th percentile is around three times the 99.5th percentile and the maximum is up to almost five times the 99.9th percentile, which shows that the dataset has heavy tails, as can be also seen in the box-plots of the logarithm of the yearly expenses in Figure \ref{box-plot}. 

\begin{table}
	\small
	\centering
	\caption{Descriptive statistics of the yearly expenses. The percentiles, mean and standard deviation refer to the truncated data values, i.e., consider only the yearly expenses which are greater than R\$ 100.}
	\label{descriptive}
	\begin{tabular}{lllllll}
		\hline
		Year & & 2006 & 2007 & 2008 & 2009 & All data \\
		\hline 
		\multicolumn{2}{l}{Size} & 70,186 & 71,814 & 73,038 & 74,418 & 289,456 \\ 
		\multicolumn{2}{l}{Percentage of $<$ R\$ 100} & 60 & 81 & 82 & 51 & 69 \\ 
		\hline
		& 25 & 195 & 151 & 154 & 249 & 190 \\ 
		& 50 & 367 & 237 & 259 & 514 & 366 \\ 
		& 75 & 807 & 453 & 510 & 1,147 & 830 \\ 
		& 90 & 1,901 & 1,033 & 1,180 & 2,782 & 2,040 \\ 
		& 95 & 3,610 & 2,309 & 2,737 & 5,662 & 4,145 \\ 
		Percentiles & 96 & 4,412 & 3,035 & 3,632 & 7,123 & 5,186 \\ 
		& 97 & 5,756 & 4,227 & 4,950 & 9,593 & 6,878 \\ 
		& 98 & 8,337 & 6,633 & 7,906 & 14,754 & 10,629 \\ 
		& 99 & 15,971 & 13,168 & 15,583 & 27,675 & 20,023 \\ 
		& 99.5 & 28,700 & 22,990 & 29,614 & 50,549 & 35,821 \\ 
		& 99.9 & 106,744 & 61,111 & 68,552 & 151,114 & 116,643 \\ 
		\hline 
		\multicolumn{2}{l}{Maximum} & 377,862 & 295,736 & 279,450 & 675,440 & 675,440 \\ 
		\multicolumn{2}{l}{Mean} & 1,313 & 870 & 991 & 2,028 & 1,485 \\ 
		\multicolumn{2}{l}{Standard Deviation} & 7,082 & 4,689 & 4,898 & 10,966 & 8,387 \\  
		\hline
	\end{tabular}
\end{table}

\begin{figure}[t]
	\centering
	\includegraphics[width = 0.75\linewidth, height=0.2\textheight]{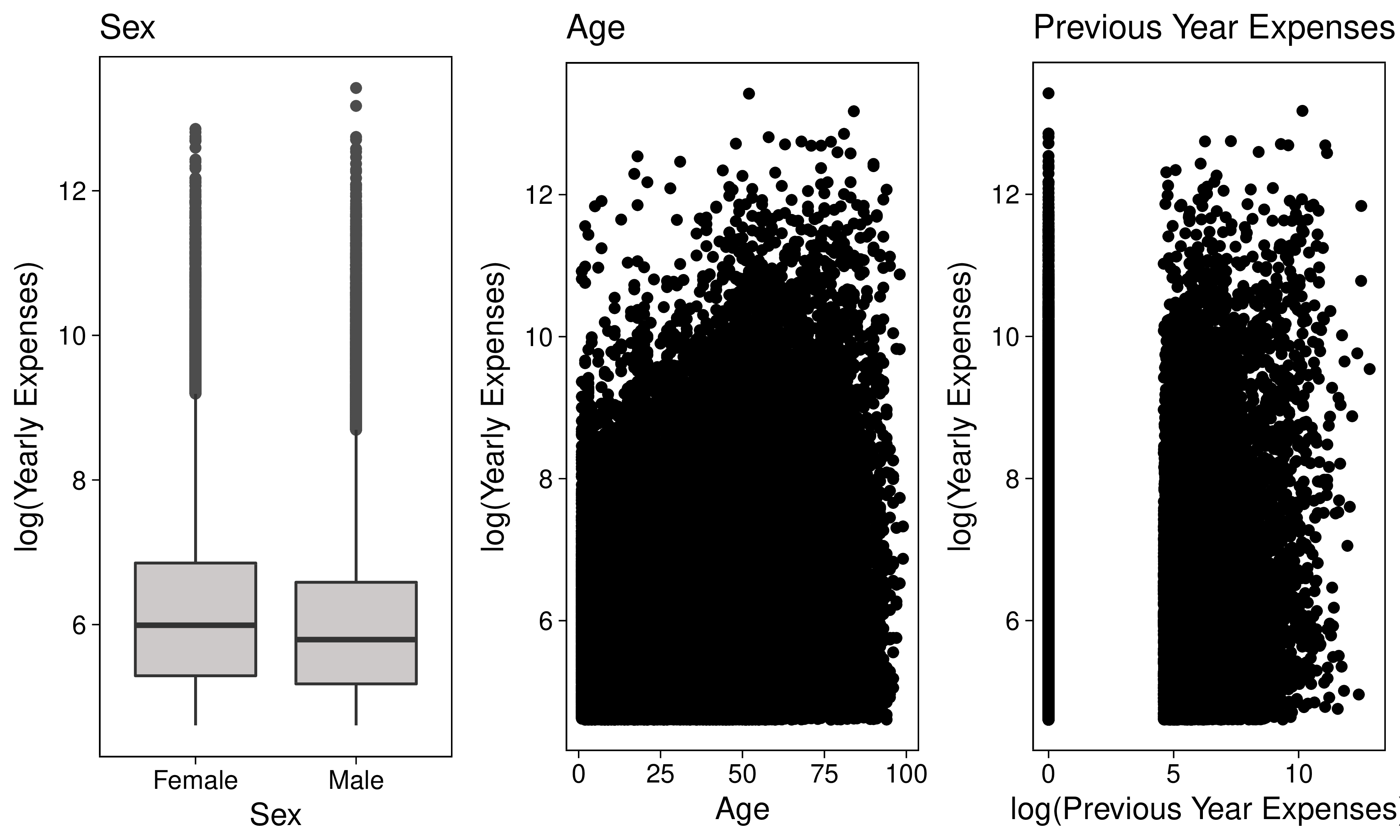}
	\caption{Dispersion of the logarithm of the yearly expenses by each one of the covariates. The yearly expenses lesser than R\$ 100 were omitted for a better visualization.}
	\label{explanatory}
\end{figure}

Figure \ref{explanatory} shows the dispersion of the logarithm of the yearly expenses by each of the covariates that are considered on the regression model, i.e., age, sex and the logarithm of the previous year expenses. The data considered for the regression model contemplate the yearly expenses of 2007, 2008 and 2009, and regards only patients that were enrolled in the insurance program in the considered year and in the previous year, which amounts to 214,925 observations. Figure \ref{explanatory} does not yield any clear relation between the logarithm of the yearly expenses and age or previous year expenses, although it seems that women tend to have greater yearly expenses than men.

\vskip 4mm

\noindent 6.2 \zgld MODEL FIT

We first fit \zgld and HGPD curves to the yearly expenses of each year (2006, 2007, 2008 and 2009) without considering any covariate. All models are fitted to the logarithm of the yearly expenses in order to obtain better fitted models and for computation optimization, as the non-transformed data has some extreme outliers, which makes it hard to fit a model properly. The goodness-of-fit is established graphically by the use of QQ-plots and the histogram of the data superimposed by the estimated curves. The fitted curves are also compared with the kernel density estimate in order to establish which is the model that best fit the data objectively. See \citet{bickel1973} and \citet{fan1994} for examples of how the kernel density estimation is used for assessing goodness-of-fit.  We apply the method proposed by \citet{sheather1991} in order to choose the bandwidth of the kernel estimate, and we choose the probability density function of a standard normal distribution as the kernel. For more details on kernel estimation see \citet{silverman1986}.

In order to compare the fitted curves to the kernel density estimate we use three different distance measures: the global distance, the $L^{2}$ norm and the $L^{\infty}$ norm that are given respectively by $D(\hat{f},\hat{f}_{k})  = \frac{1}{n} \sum_{i=1}^{n} [\hat{f}(y_{i}) - \hat{f}_{k}(y_{i})]^{2}$, $\lVert \hat{f} - \hat{f}_{k} \lVert_{2} =  \Big[ \int_{0}^{+ \infty} \big|\hat{f}(y) - \hat{f}_{k}(y)\big|^{2} dy \Big]^{1/2}$ and $\lVert \hat{f} - \hat{f}_{k} \lVert_{\infty} =  \max\limits_{y \in \mathbb{R}^{+}} |\hat{f}(y) - \hat{f}_{k}(y)|$, in which $\hat{f}$ is the parametric curve (\gld or \gpds) fitted to the non-zero yearly expenses and $f_{k}$ is the kernel density estimate. Note that the probability mass at zero is the same for all fitted curves, so there is no need to compare them regarding the zero valued yearly expenses.

\begin{figure}[t]
	\centering
	\includegraphics[width = 0.75\linewidth, height=0.2\textheight]{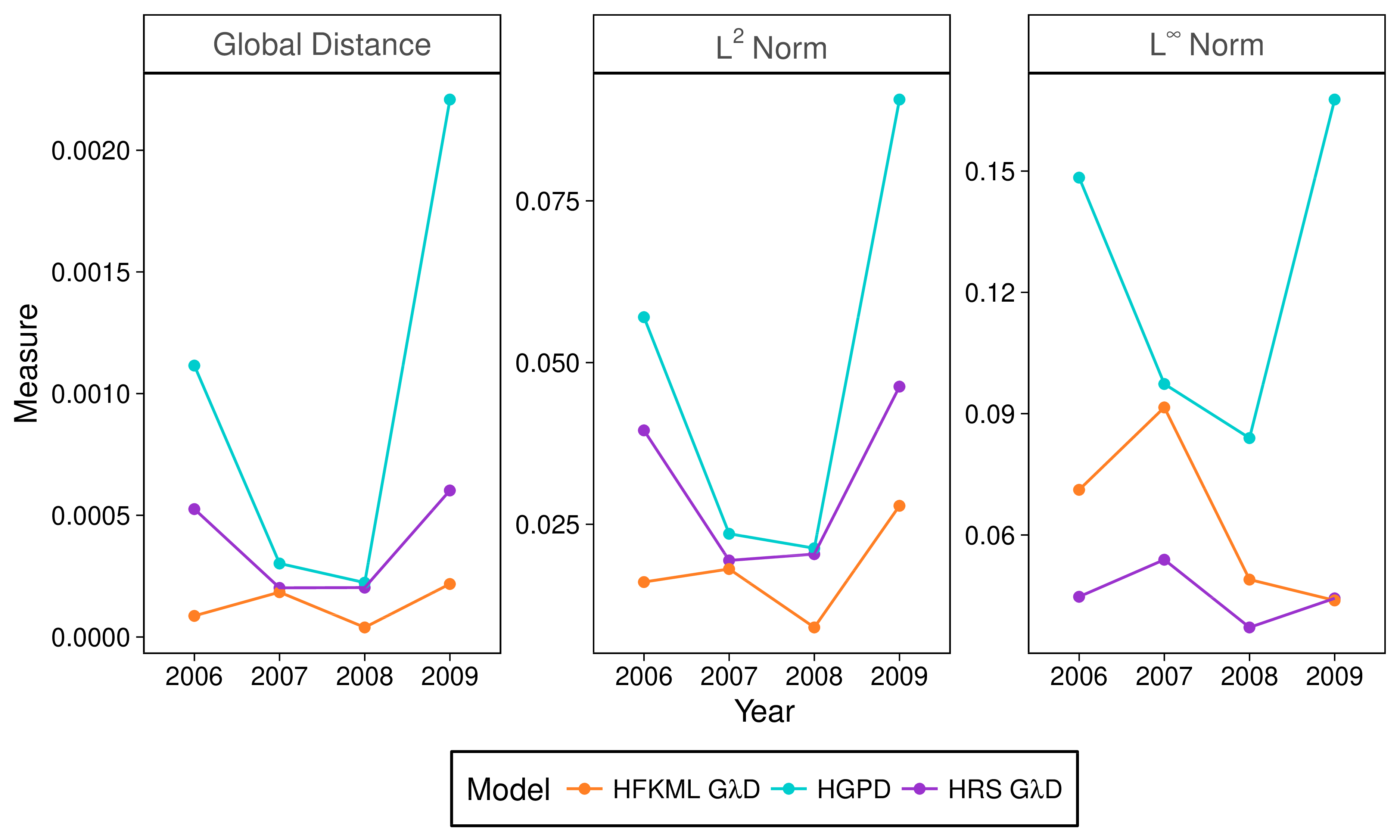}
	\caption{The distance between the fitted curve and the kernel density estimate for each model and year.}
	\label{distancesM}
\end{figure}

\begin{figure}
	\begin{flushleft}
		\noindent DIAGNOSTIC PLOTS
	\end{flushleft}
	\centering
	\includegraphics[height=0.93\textheight]{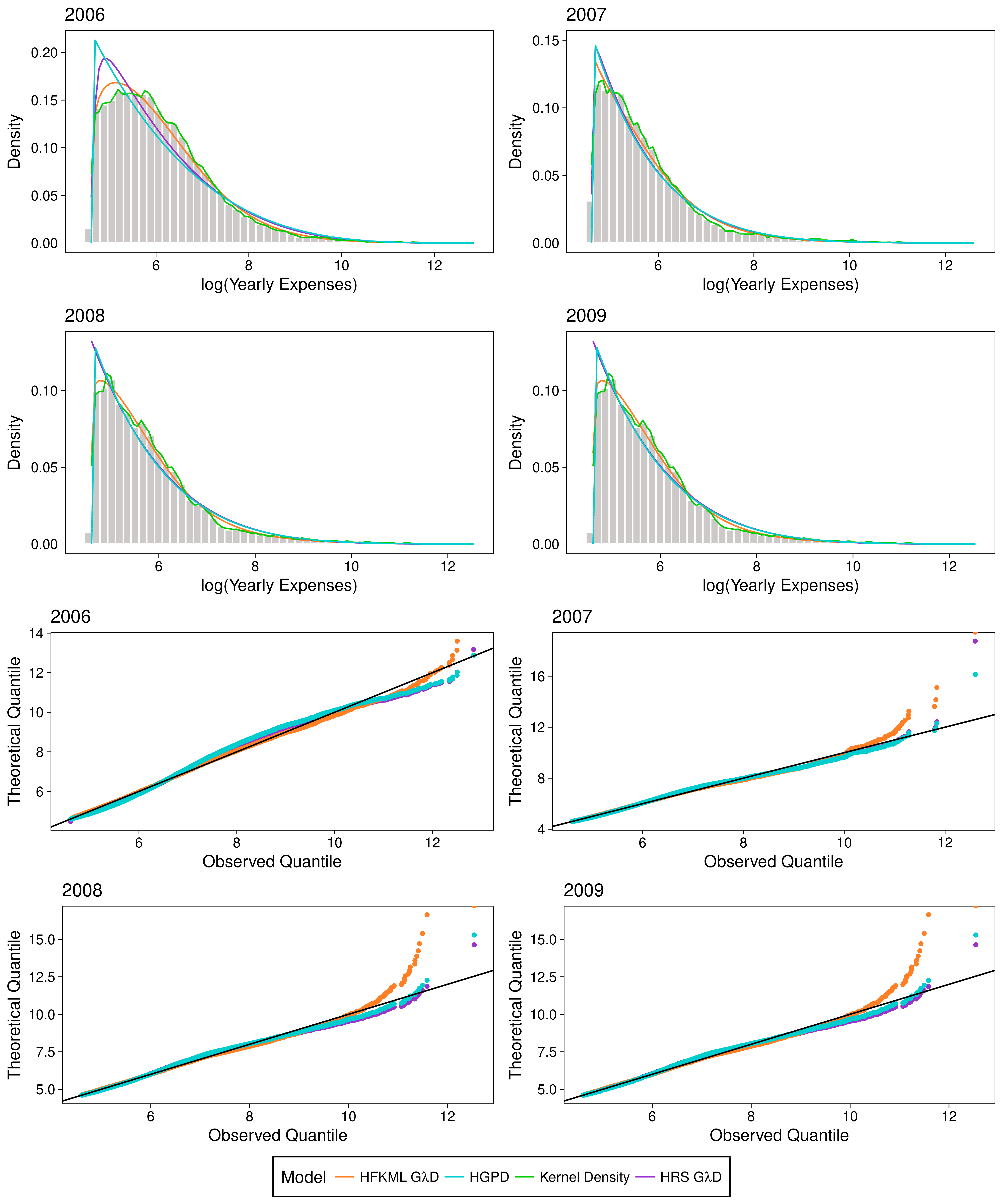}
	\caption{\footnotesize The top four plots display the histogram of the data superimposed by the estimated curves of the \zgld and HGPD models, for each year. The bottom four plots display the QQ-plot between the sample quantiles and the theoretical quantiles of the \zgld and HGPD models, for each year.}
	\label{HQQano}
\end{figure}

The estimated parameters for each year and model are displayed in Table \ref{paramFit}. The estimated parameters differ significantly from one year to another, for all fitted models, although we observe in every year that the fitted G$\lambda$Ds are highly skewed, as the values of $\lambda_{1}$ and $\lambda_{2}$ are quite different. In Figure \ref{distancesM} we see that the densities estimated by the HRS and \zfgld are closer to the kernel estimate density for all years, by all distance measures. Furthermore, Figure \ref{HQQano} displays the histogram of the logarithm of the yearly expenses superimposed by the fitted curves of the \zgld and HGPD models and the QQ-plots between the empirical and fitted distributions, for all years, from which it can be seen that the \zgld models fit the data best for low values (near the threshold 4.61), and that the \zrgld and HGPD models fit as good the tail, while the \zfgld model seems to fit it poorer. 

\begin{table}[t]
	\small
	\centering
	\caption{Estimated parameters for the \zgld and HGPD models fitted to the yearly expenses, for each year.}
	\label{paramFit}
	\begin{tabular}{rrlrrrrrrr}
		%\hiderowcolors
		\hline
		\multirow{2}{*}{Year} & \multirow{2}{*}{$\lambda_{0}$} & & & \gld & & & & \gpd & \\ \cline{3-10}
		& & Par & $\lambda_{1}$ & $\lambda_{2}$ & $\lambda_{3}$ & $\lambda_{4}$ & Scale & Shape & Location \\ 
		\hline
		\multirow{2}{*}{2006} & \multirow{2}{*}{0.60} & RS & 4.74 & 0.12 & 0.0032 & 0.20 & \multirow{2}{*}{1.80} & \multirow{2}{*}{-0.22} & \multirow{2}{*}{4.61} \\
		&  & FKML &  5.74 & 1.13 & 0.78 & 0.03 &  &  &  \\ 
		\hline
		\multirow{2}{*}{2007} & \multirow{2}{*}{0.81} & RS & 4.62 & 0.07 & 0.0002 & 0.08 & \multirow{2}{*}{1.22} & \multirow{2}{*}{-0.09} & \multirow{2}{*}{4.61} \\ 
		&  & FKML & 5.30 & 1.37 & 1.05 & -0.07 &  &  &  \\ 
		\hline
		\multirow{2}{*}{2008} & \multirow{2}{*}{0.82} & RS & 4.61 & 0.10 & 0 & 0.14 & \multirow{2}{*}{1.33} & \multirow{2}{*}{-0.12} & \multirow{2}{*}{4.61} \\ 
		&  & FKML & 5.39 & 1.43 & 0.89 & -0.10 &  &  &  \\ 
		\hline
		\multirow{2}{*}{2009} & \multirow{2}{*}{0.51} & RS & 5.20 & 0.11 & 0.02 & 0.18 & \multirow{2}{*}{2.17} & \multirow{2}{*}{-0.25} & \multirow{2}{*}{4.61} \\ 
		&  & FKML & 6.06 & 1.07 & 0.64 & 0.04 &  &  &  \\ 
		\hline
	\end{tabular}
\end{table}

From the diagnostic plots in Figure \ref{HQQano} we see that the major advantage of the \zgld over the HGPD is that it is not necessarily threshold modal and monotonically decreasing so that it fits better the \textit{bulk} of the distribution, i.e., the values near the threshold, when the distribution mode is greater than the threshold. Nevertheless, the \zrgld and the HGPD fit better the tail of the distribution, while the \zfgld fits better its bulk for it is the distribution with best overall fit according to the distance measures. Therefore, the \zgld models fit better the data, especially the \zrglds, although the HGPD fits the right tail of the distribution as good as them.

In order to chose between the proposed hurdle models, one should observe the nature of the data the models seek to fit. Indeed, although the \gpd has a highly flexible right tail, which makes it useful for fitting heavy tailed data, its left tail is not quite flexible, which makes it a poor choice for modelling data that demands flexibility in both tails. On the other hand, both tails of the \gld are flexible, so that it is a more robust choice when comparing to the \gpds. As the parametrizations of the RS and \fgld are not equivalent, one must also chose between them, what may be done by observing the quality of each fit by applying tools as the distance to the kernel estimate or diagnostic plots.

\vskip 4mm

\noindent 6.3 \zgld REGRESSION MODEL

In this section, \zgld regression models are fitted to the logarithm of the yearly expenses and compared with the HGPD GLM by the use of error and normalised quantile residuals. The estimated parameters of the logistic regression, i.e., the parameters of the model for the probability mass at zero, are the same for all the fitted models, as they are orthogonal to the parameters of the models for the non-zero values. Also, the logit modelled in the logistic regression is the logit of the expense being less than R\$ 100, as the yearly expenses were truncated at R\$ 100. In order to fit the models, we assume that, given the logarithm of the previous year expenses, the age and the sex, the logarithm of the yearly expenses are independent, even the expenses that refer to the same person in different years, so that we have a sample of the model variables. 

The estimated parameters of the logistic regression for the zero-valued data are presented in Table \ref{logistic}, in which the contrast used for the sex is ``treatment'' in which the female sex is the base. The minus sign of the estimated coefficient of the logarithm of the previous year expenses and the age shows that as greater the previous year expense or the age of a person, the lesser is the probability of him having less than R\$ 100 in yearly expenses, while the plus sign of the estimated coefficient for the male sex shows that men are more likely to have yearly healthcare expenses lesser than R\$ 100 than women.

\begin{table}[h]
	\small
	\centering
	\caption{Estimated parameters of the logistic regression that models the logit of the yearly expenses being less than R\$ 100.}
	\label{logistic}
	\begin{tabular}{rrrrc}
		\hline
		Parameter & Estimate & SE & t value & p-value \\
		\hline 
		Intercept & 1.6266 & 0.0121 & 134.8826 & $< 2e-16$ \\ 
		LE & -0.1253 & 0.0018 & -71.4760 & $< 2e-16$ \\ 
		Male & 0.2093 & 0.0098 & 21.2574 & $< 2e-16$ \\ 
		Age & -0.0159 & 0.0002 & -63.9644 & $< 2e-16$ \\ 
		\hline
		\multicolumn{5}{l}{\footnotesize SE: Standard Deviation; LE: Logarithm of the previous year expenses} \\
	\end{tabular}
\end{table}

The estimated parameters of both parametrizations of the \gld regression and of the GPD GLM for the non-zero data values are presented in Tables \ref{non-zero} and \ref{non-zero2}, in which the female sex is again taken as the base for the ``treatment'' contrast of sex. On the one hand, as the zero is in the 99\% confidence interval for all covariate's coefficients of the \zfgld model, there is no evidence that the location of the distribution depends on any of the covariates at a significance of 1\% and we may regard these parameters as zero. On the other hand, all the parameters of the \zrgld and HGPD model are different of zero at a significance of 1\%, so that we regard only the estimated coefficients of these models.

\begin{table}[h]
	\small
	\centering
	\caption{Estimated parameters and numerical confidence intervals of the \zrgld and \zfgld models.}
	\label{non-zero}
	\begin{tabular}{crccc}
		\hline
		\multirow{2}{*}{Parametrization} & \multirow{2}{*}{Parameter} & \multirow{2}{*}{Estimate} & \multicolumn{2}{c}{Confidence Interval} \\ \cline{4-5}
		& & & L. B. (0.5\%) & U. B. (99.5\%) \\ 
		\hline
		\multirow{8}{*}{\zfgld}& Intercept & 6.13 & 6.11 & 6.24 \\ 
		& LE & -0.0000215 & -0.0050842 & 0.0016275 \\
		& Male &  -0.0003554 & -0.2338218 & 0.0392448 \\ 
		& Age & 0.0000259 & -0.0002445 & 0.0003319 \\
		& $\lambda_{1}$ & -0.41 & -0.59 & -0.29 \\ 
		& $\lambda_{2}$ & 1.07 & 0.94 & 1.56 \\ 
		& $\lambda_{3}$ & 0.84 & 0.50 & 1.02 \\ 
		& $\lambda_{4}$ & 0.02 & -0.26 & 0.09 \\ 
		\hline
		\multirow{8}{*}{\zrgld}& Intercept & 6.10 & 6.08 & 6.11 \\
		& LE & 0.0013937 & 0.0005463 & 0.0023140 \\ 
		& Male & -0.0126310 & -0.0182945 & -0.0074669 \\ 
		& Age & 0.0009363 & 0.0007947 & 0.0010634 \\ 
		& $\lambda_{1}$ & -1.41 & -1.43 & -1.40 \\ 
		& $\lambda_{2}$ & 0.1102 & 0.1061 & 0.1142 \\  
		& $\lambda_{3}$ & 0.0023749 & 0.0021813 & 0.0025770 \\ 
		& $\lambda_{4}$ & 0.19 & 0.18 & 0.20 \\ 
		\hline
		\multicolumn{5}{l}{\footnotesize LE: Logarithm of the previous year expenses; L. B.: Lower bound; U. B.: Upper bound}
	\end{tabular}
\end{table}

\begin{table}[h]
	\small
	\centering
	\caption{Estimated parameters and p-values of the HGPD model.}
	\label{non-zero2}
	\begin{tabular}{rccc}
		\hline
		Parameter & Estimate & SE & p-value \\ 
		\hline
		Shape &  0.9924 & 2.58e-13 & $< 2e-16$ \\ 
		Intercept & 4.6576 & 0.0091 & $< 2e-16$ \\ 
		LE & 0.0152 & 0.0021 & $8.26e-13$ \\ 
		Male & -0.1198 & 0.0125 & $< 2e-16$ \\ 
		Age & 0.0082 & 0.0003 & $< 2e-16$ \\ 
		\hline
		\multicolumn{4}{l}{\footnotesize SE: Standard Deviation; LE: Logarithm of the previous year expenses.}
	\end{tabular}
\end{table}

The signs of the estimated parameters of the \zrgld and HGPD models are exchanged when comparing with the signs of the parameters in Table \ref{logistic}, which is consistent. Indeed, we see that as greater the previous year expenses or the age, the greater is the location parameter of the \zgld and the mean of the HGPD, and that the location parameter (and mean) of the male sex is lesser than the female's, confirming what were observed in the box-plot in Figure \ref{explanatory}. Therefore, we obtain the same kind of interpretation for the yearly expenses from both the logistic regression, \zrgld model and the HGPD GLM: as greater the previous year expenses or the age, the greater the expense; and women have greater expense than men. 

\begin{figure}[h]
	\centering
	\includegraphics[height=0.98\textheight]{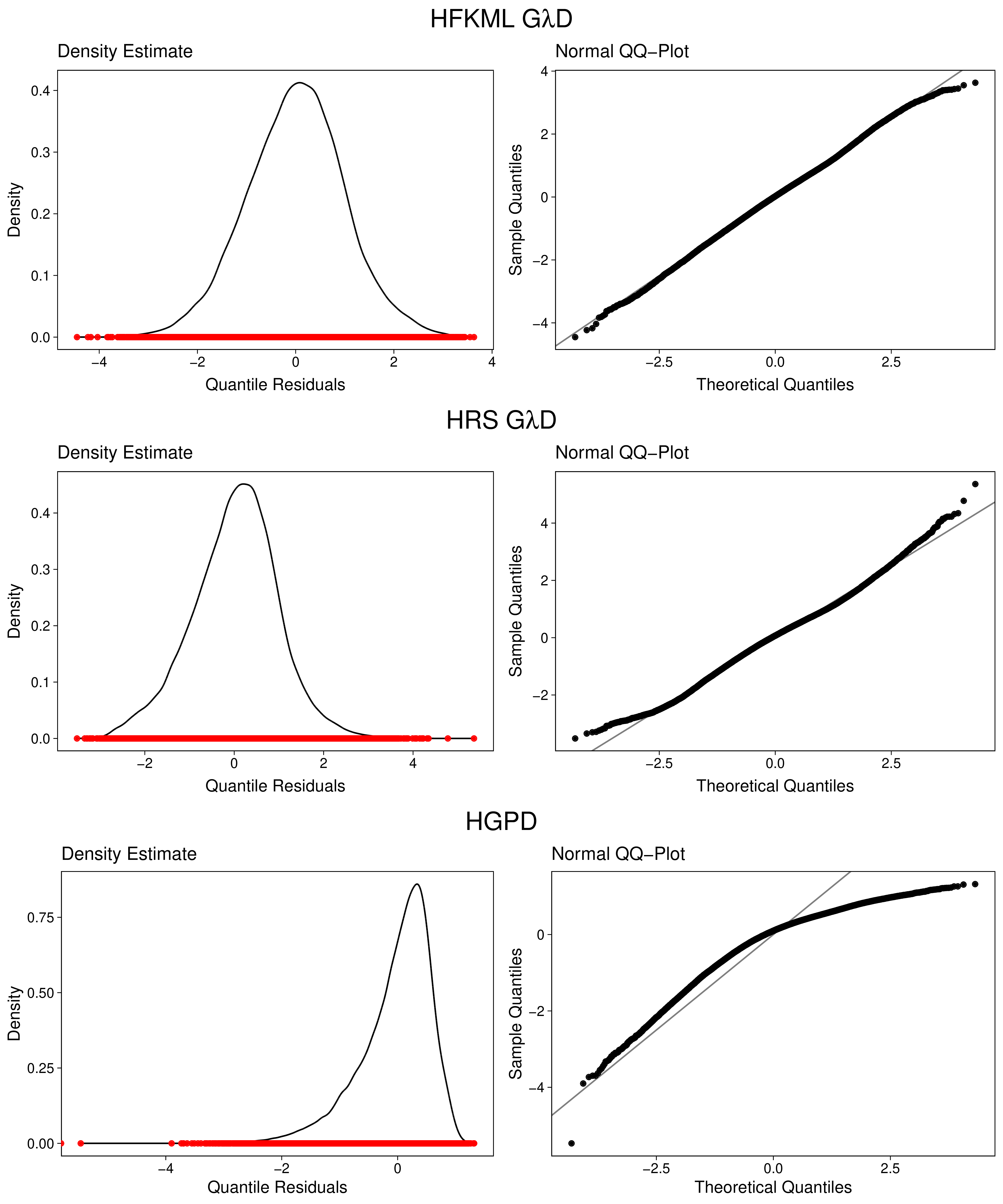}
	\caption{\footnotesize Estimated density and Normal QQ-plot of the normalised quantile residuals of the \zrglds and \zfgld regressions models and the HGD GLM for the non-zero yearly expenses.}
	\label{qdiagGLD}
\end{figure}

\begin{figure}[h]
	\centering
	\includegraphics[height=0.97\textheight]{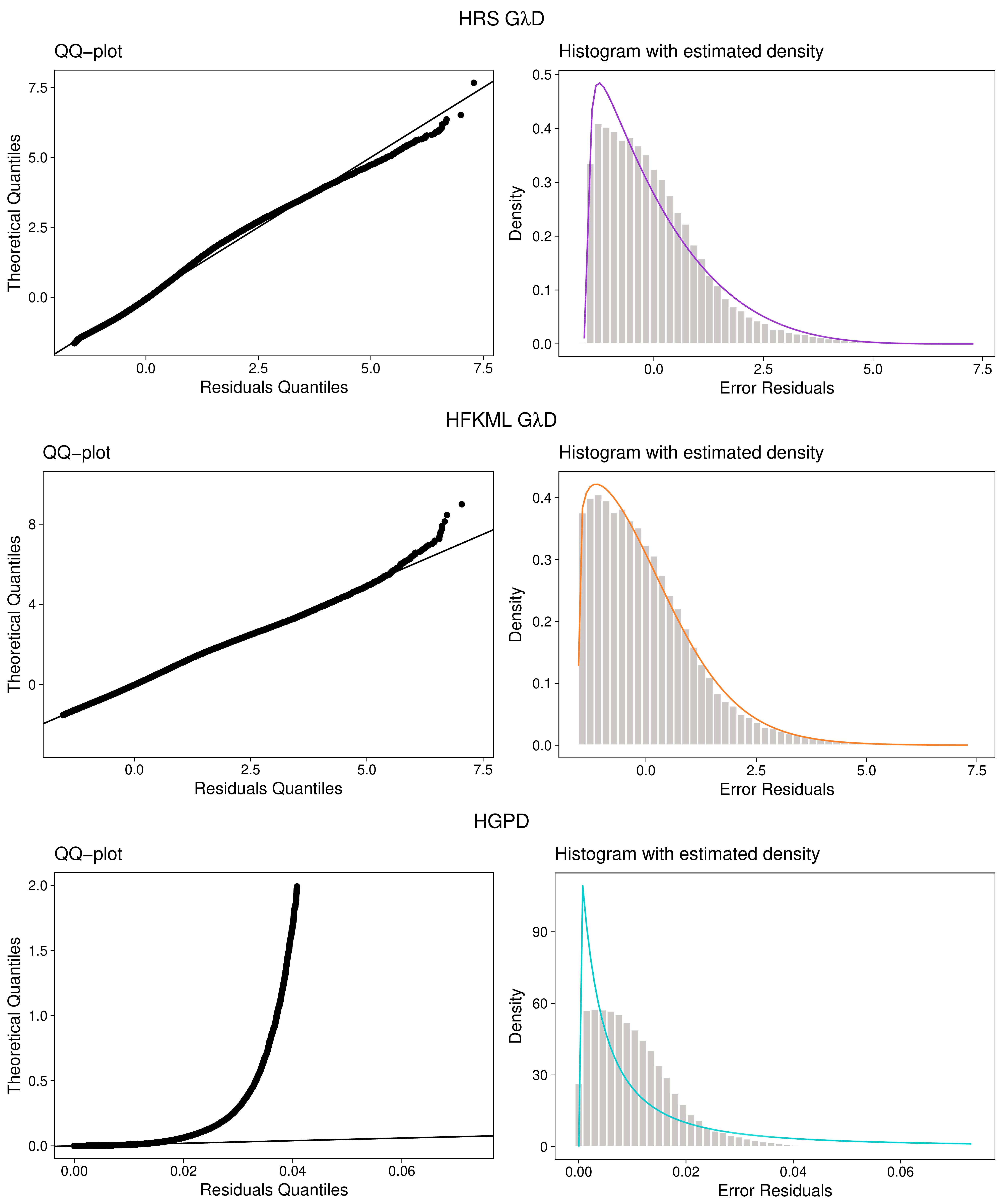}
	\caption{\footnotesize Diagnostics for the \zrglds and \zfgld regression models and the HGPD GLM for the non-zero yearly expenses. The histograms are that of the respective error residuals and are superimposed by their theoretical distribution. The QQ-plots compare the empirical quantiles of the error residuals with their theoretical quantiles.}
	\label{diagGLD}
\end{figure}

The diagnostic plots for the \zgld models and the HGPD GLM are presented in Figures \ref{qdiagGLD} and \ref{diagGLD}. Figure \ref{qdiagGLD} display plots of the normalized quantile residuals, while Figure \ref{diagGLD} display plots of the error residuals. Figure \ref{qdiagGLD} yields that the \zrgld and \zfgld models are fairly fitted, as the distributions of their normalized quantile residuals do not greatly deviate from the normal distribution. Furthermore, from Figure \ref{diagGLD} it may be established that the \zrgld and \zfgld models are well-fitted, as the points of its error residuals QQ-plot are distributed around the line with intercept zero and slope one. In fact, when comparing with the HGPD GLM, the \zrgld and \zfgld models seem to better fit the data. 

On the other hand, the fit of the HGPD GLM is not good, as its error residuals do not seem to be distributed as a GPD and its normalised quantile residuals are highly skewed. The HGPD GLM does not properly fit the residuals because the data is not threshold modal and the fitted distribution is supposed to have infinity expectation, as can be seem from the estimate of the shape parameter that is close to one. The lack of flexibility of its left tail makes the GPD improper to fit data that presents a behaviour on the left tail that is not threshold modal and monotonically decreasing. Furthermore, the GLM framework is restricted to GPDs that have finite expectation, i.e., such that $\xi < 1$. On the other hand, the \gld is exactly the opposite of the \gpd in the matter of tail flexibility, as its tails may have different shapes. Moreover, the \zgld models the location of the distribution, so that it may fit distributions with infinite expectation. 

In general, when choosing between the proposed hurdle models, one must take into account the statistical significance of its parameters, and carefully analyse the behaviour of the normalized quantile and error residuals. The \gld regression models are more robust, as are also adequate when the conditional distribution of the response variable given the covariate has infinite mean or is not monotonically decreasing with the threshold as the mode. Nevertheless, one has also to choose between the RS and \fglds, which are not equivalent models and, in order to do so, must carefully analyse both models, and choose the one that best fulfils the objective of the regression model, e.g., best predicts an outcome or best fit the dataset. 

\begin{figure}[t]
	\centering
	\includegraphics[width = 0.75\linewidth, height=0.2\textheight]{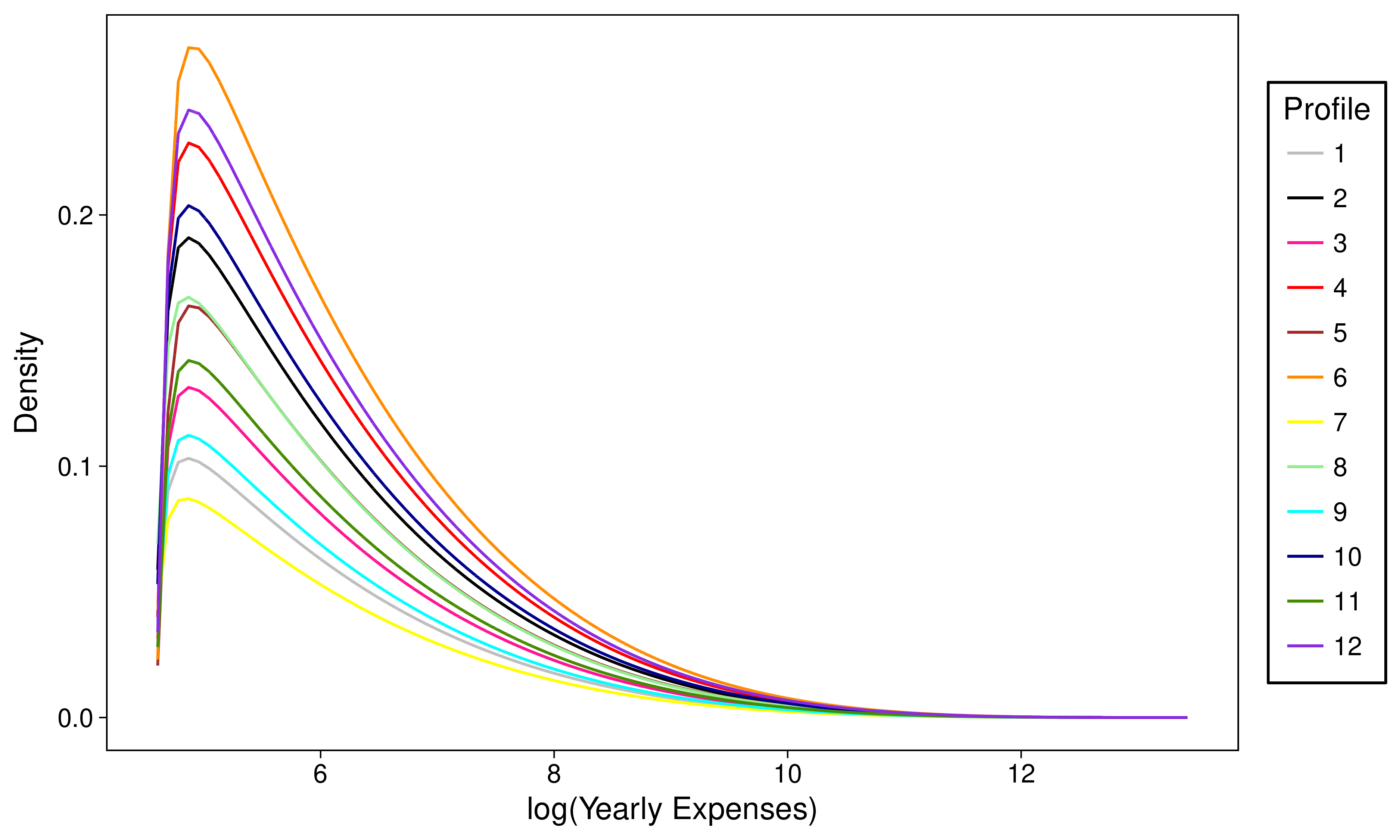}
	\caption{Estimated curves for the profiles in Table \ref{percentiles} given by the \zrgld model.}
	\label{Fprofile}
\end{figure}

An interesting feature of the \zgld regression models is that the fitted curve takes into account the probability mass at zero, so that we may readily see what are the profiles, i.e., combinations of the covariate's levels, that tend to have great and low expenses. As an example, we consider 12 profiles, that are presented in Table \ref{percentiles} and whose \zrgld fitted curves are displayed in Figure \ref{Fprofile}. On the one hand, the location of the curves is almost the same for all profiles, even though there are profiles that differ reasonably on all the covariates. On the other hand, the probability mass at zero differs significantly from one profile to another, as can be seen from the area under each curve, that represents one minus the probability mass at zero. The exponential of selected percentiles for the 12 profiles are presented in Table \ref{percentiles}, in which we observe that the percentiles differ significantly from one profile to another and their values are a reflex of the estimated parameters of Tables \ref{logistic} and \ref{non-zero}.

\begin{table}[h]
	\small
	\centering
	\caption{The covariates of each profile, their location, $\lambda_{0}$ and selected estimated percentiles for the yearly expenses from the \zrgld model. The location and percentiles are exponentiated.}
	\label{percentiles} 
	\resizebox{\linewidth}{!}{
	\begin{tabular}{cccccccccccc}
		\hline
		\multirow{2}{*}{Profile} & \multirow{2}{*}{Age} & \multirow{2}{*}{Sex} & \multirow{2}{*}{LE} & \multirow{2}{*}{$\lambda_{0}$} & \multirow{2}{*}{Location} & \multicolumn{6}{c}{Selected Percentiles} \\ \cline{7-12}
		& & & & & & 75th & 90th & 95th & 99th & 99.5th & 99.9th \\ 
		\hline
		1 & 20 & F & 0 & 0.79 & 452.53 & 0 & 360.78 & 948.72 & 5768.93 & 10763.16 & 34689.11 \\ 
		2 & 20 & F & 7 & 0.61 & 456.97 & 228.98 & 865.29 & 2040.61 & 10176.05 & 17736.15 & 50318.16 \\  
		3 & 40 & F & 0 & 0.73 & 461.08 & 122.14 & 522.95 & 1314.88 & 7374.60 & 13383.05 & 40952.71 \\ 
		4 & 40 & F & 7 & 0.53 & 465.60 & 309.58 & 1115.03 & 2554.16 & 12069.66 & 20650.44 & 56586.22 \\ 
		5 & 60 & F & 0 & 0.66 & 469.80 & 183.74 & 725.48 & 1754.89 & 9169.98 & 16241.66 & 47487.74 \\ 
		6 & 60 & F & 7 & 0.45 & 474.40 & 398.82 & 1381.68 & 3089.37 & 13958.85 & 23516.69 & 62600.83 \\ 
		7 & 20 & M & 0 & 0.82 & 446.85 & 0 & 275.79 & 749.75 & 4834.56 & 9202.52 & 30793.22 \\ 
		8 & 20 & M & 7 & 0.65 & 451.23 & 182.43 & 716.00 & 1726.10 & 8963.52 & 15842.01 & 46133.72 \\ 
		9 & 40 & M & 0 & 0.77 & 455.30 & 0 & 411.71 & 1065.33 & 6293.60 & 11626.53 & 36784.42 \\ 
		10 & 40 & M & 7 & 0.58 & 459.76 & 255.32 & 947.96 & 2212.09 & 10817.96 & 18728.74 & 52469.37 \\ 
		11 & 60 & M & 0 & 0.71 & 463.90 & 141.94 & 587.78 & 1457.46 & 7969.18 & 14336.51 & 43159.53 \\ 
		12 & 60 & M & 7 & 0.50 & 468.45 & 339.34 & 1204.89 & 2735.73 & 12717.94 & 21637.32 & 58667.27 \\ 
		\hline
		\multicolumn{12}{l}{\footnotesize LE: Logarithm of the previous year expenses.}
	\end{tabular}}
\end{table}

\vskip 4mm

\noindent FINAL REMARKS

The \zgld models proposed in this paper have a great potential for applications, not only to healthcare expenses data, but also to any highly skewed data, with excess of zeros and heavy tails. According to the results obtained in Section 6, we may argue that the HGPD is in general as good as the \zgld when fitting  unimodal monotonically decreasing distributions, while the \zgld seems to better fit data that demands a higher flexibility in its left tail. Therefore, the methods developed in this paper bring contributions to the state-of-the-art in modelling heavy tailed clumped-at-zero data.

Although the \zgld fits best some kinds of data, it is still necessary to improve its methods of estimation, especially what concerns the asymptotic properties of the estimators and the computation of the estimates, that may take days, depending on the size of the data and the number of parameters. Therefore, a more theoretical research about the \zgld and the optimization of the algorithms used to estimate its parameters are interesting topics for future researches. 
\vskip 4mm

\noindent ACKNOWLEDGEMENTS

We would like to thank \textit{Sabesprev} who kindly provided the dataset used in this paper.
\vskip 4mm

\FloatBarrier
\noindent SUPPLEMENTARY MATERIAL

The data analysis has been performed in the 3.4.2 version of \textbf{R} \citep{R} by the adaptation of functions of the \textbf{GAMLSS} \citep{gamlss}, \textbf{GLDEX} \citep{su2007b} and \textbf{GLDReg} \citep{su2016} packages. In the on-line supplementary material we provide an \textbf{R} package with functions to fit all the models of this paper and an \textbf{R} script that reproduce all tables and figures of this paper.
 
\vskip 4mm

%\printendnotes

% Submissions are not required to reflect the precise reference formatting of the journal (use of italics, bold etc.), however it is important that all key elements of each reference are included.
\bibliographystyle{apa}
\bibliography{ref.bib}
\vskip 4mm

\newpage
\FloatBarrier

\end{document}